\documentclass[11pt]{article}
\usepackage{amssymb,amsmath,amsfonts}
\usepackage{graphicx}
\usepackage{graphics}
\usepackage{eepic,epsfig}

\@addtoreset{equation}{section}

\makeatother

\begin{document}

\title{Casimir densities for a spherical boundary \\
in de Sitter spacetime}
\author{K. Milton$^{1}$, A.~A. Saharian$^{2}$ \\
\\
\textit{$^{1}$Homer L. Dodge Department of Physics and Astronomy, }\\
\textit{University of Oklahoma, Norman, Oklahoma 73019-2061, USA }\vspace{%
0.3cm}\\
$^{2}$\textit{Department of Physics, Yerevan State University,}\\
\textit{1 Alex Manoogian Street, 0025 Yerevan, Armenia}}
\maketitle

\begin{abstract}
Two-point functions, mean-squared fluctuations, and the vacuum expectation
value of the energy-momentum tensor operator are investigated for a massive
scalar field with an arbitrary curvature coupling parameter, subject to a
spherical boundary in the background of de Sitter spacetime. The field is
prepared in the Bunch-Davies vacuum state and is constrained to satisfy
Robin boundary conditions on the sphere. Both the interior and exterior
regions are considered. For the calculation in the interior region, a
mode-summation method is employed, supplemented with a variant of the
generalized Abel-Plana formula. This allows us to explicitly extract the
contributions to the expectation values which come from de Sitter spacetime
without boundaries. We show that the vacuum energy-momentum tensor is
non-diagonal with the off-diagonal component corresponding to the energy
flux along the radial direction. With dependence on the boundary condition
and the mass of the field, this flux can be either positive or negative.
Several limiting cases of interest are then studied. In terms of the
curvature coupling parameter and the mass of the field, two very different
regimes are realized, which exhibit monotonic and oscillatory behavior of
the vacuum expectation values, respectively, far from the sphere. The decay
of the boundary induced expectation values at large distances from the
sphere is shown to be power-law (monotonic or oscillating), independent of
the value of the field mass. The expressions for the Casimir densities in
the exterior region are generalized for a more general class of
spherically-symmetric spacetimes inside the sphere.
\end{abstract}

\bigskip


\bigskip

\section{Introduction}

De Sitter (dS) spacetime is one of the simplest and most interesting
spacetimes allowed by general relativity. Quantum field theory in this
background has been extensively studied during the past two decades. Much of
the early interest was motivated by the questions related to the
quantization of fields on curved backgrounds. dS spacetime has a high degree
of symmetry and numerous physical problems are exactly solvable on this
background. The importance of this theoretical work increased by the
appearance of the inflationary cosmology scenario \cite{Lind90}. In most
inflationary models, an approximately dS spacetime is employed to solve a
number of problems in standard cosmology. During an inflationary epoch,
quantum fluctuations in the inflaton field introduce inhomogeneities which
play a central role in the generation of cosmic structures from inflation.
More recently, astronomical observations of high redshift supernovae, galaxy
clusters, and cosmic microwave background \cite{Ries07} indicate that at the
present epoch the universe is accelerating and can be well approximated by a
world with a positive cosmological constant. If the universe would
accelerate indefinitely, the standard cosmology would lead to an asymptotic
dS universe. Hence, the investigation of physical effects in dS spacetime is
important for understanding both the early universe and its future. Another
motivation for investigations of dS-based quantum theories is related to the
holographic duality between quantum gravity on dS spacetime and a quantum
field theory living on a boundary identified with the timelike infinity of
dS spacetime \cite{Stro01}.

In dS spacetime, the interaction of fluctuating quantum fields with the
background gravitational field gives rise to vacuum polarization. The
Casimir effect presents another type of vacuum polarization induced by the
presence of boundaries. This effect is among the most striking macroscopic
manifestations of non-trivial properties of the quantum vacuum. It has
important implications on all scales, from subnuclear to cosmological. The
reflecting boundaries alter the zero-point modes of a quantized field and
shift the vacuum expectation values of quantities, such as the energy
density and stresses. As a result, forces arise acting on constraining
boundaries. The particular features of these forces depend on the nature of
the quantum field, the type of spacetime manifold, the boundary geometry,
and the specific boundary conditions imposed on the field. Since the
original work by Casimir many theoretical and experimental works have been
done on this problem (see, e.g., Ref.~\cite{Most97} and references therein).

In the present paper we consider a problem with both types of vacuum
polarization. Namely, we evaluate the vacuum expectation values for the
field squared and the energy-momentum tensor of a scalar field with general
curvature coupling parameter induced by a spherical boundary on the
background of $(D+1)$-dimensional dS spacetime. Historically, the
investigation of the Casimir effect for a spherical shell was motivated by
the Casimir semiclassical model of an electron. In this model Casimir
suggested \cite{Casimir53} that Poincar\'{e} stress, to stabilize the
charged particle, could arise from vacuum quantum fluctuations and the fine
structure constant could be determined by a balance between the Casimir
force (assumed attractive) and the Coulomb repulsion. However, as has been
shown by Boyer \cite{Boyer}, the Casimir energy for a perfectly conducting
sphere is positive, implying a repulsive force. This result was later
reconsidered by a number of authors \cite{DaviesSph}. More recently new
methods have been developed for the investigation of the Casimir effect
including direct mode summation techniques and the zeta function
regularization scheme \cite{Rome95}, semiclassical methods \cite{Scha98} in
the framework of the Gutzwiller trace formula, the optical approach \cite%
{Jaff04}, worldline numerics \cite{Gies03}, the path integral approach \cite%
{Bord85}, methods based on scattering theory \cite{Gene03}, numerical
methods based on evaluation of the stress tensor via the
fluctuation-dissipation theorem \cite{Rodr07} (for a review see Refs.~\cite%
{Most97,Lect11}). The Casimir effect for a spherical shell in an arbitrary
number of dimensions is analyzed in Refs.~\cite{MiltonSc,MiltonVec} for a
massless scalar field satisfying Dirichlet and a special type of Robin
(corresponding to the electromagnetic TM modes) boundary conditions using
the Green's function method and in Refs.~\cite{Cognola,Teo10} for the
electromagnetic field and massless scalar and spinor fields with various
boundary conditions on the basis of the zeta regularization technique. The
case for a massive vector field has been discussed in Ref.~\cite{Teo11}.

The investigation of the energy distribution inside a perfectly reflecting
spherical shell was made in Ref.~\cite{Olaussen1}. The distribution of the
other components for the energy-momentum tensor of the electromagnetic field
inside and outside a shell and in the region between two concentric
spherical shells is studied in Refs.~\cite{Brevik1,Grig1} (see also Ref.~%
\cite{Sahrev}). The investigation of the electromagnetic energy density near
a conducting boundary was first carried out by DeWitt \cite{dewitt}. The
vacuum expectation values for the energy-momentum tensor of a massive scalar
field with general curvature coupling parameter and obeying the Robin
boundary condition on spherically symmetric boundaries in $(D+1)$%
-dimensional spacetime are investigated in Ref.~\cite{Saha01}. The Casimir
densities for spherical boundaries in the background of global monopole and
Rindler-like spacetimes have been discussed in Refs.~\cite{GlobMon} and \cite%
{RindSph} respectively.

Previously, the Casimir stresses for spherical boundaries on the background
of dS spacetime have been investigated in Ref.~\cite{Seta01} for a
conformally coupled massless scalar field. In this last case the problem is
conformally related to the corresponding problem in Minkowski spacetime and
the vacuum characteristics are generated from those for the Minkowski
counterpart, just by multiplying with the conformal factor. As it has been
shown in Refs.~\cite{Saha09} for the geometry of flat boundaries,
qualitatively new features arise in the case of non-conformally coupled
fields (for the case of a minimally coupled massless field see Ref.~\cite%
{Burd11}). (For flat backgrounds, see also Ref.~\cite{miltonrev10}.) The
curvature of the background spacetime decisively influences the behavior of
boundary-induced vacuum expectation values at distances larger than the
curvature scale. Recently, the topological Casimir effect in dS spacetime
with toroidally compactified spatial dimensions has been investigated in
Ref.~\cite{Saha08}.

We have organized the paper as follows. In the next section the Wightman
function is evaluated inside a spherical boundary in dS spacetime for a
scalar field with general curvature coupling parameter and with Robin
boundary condition on the sphere. Among the most important quantities
describing the local properties of a quantum field and the corresponding
quantum back-reaction effects are the expectation values of the field
squared and of the energy-momentum tensor. These quantities in the interior
of a sphere will be investigated in Secs.~\ref{sec:phi2} and \ref{sec:EMT}.
The Wightman function for the region outside a spherical boundary is
considered in Sec.~\ref{sec:Ex}. The vacuum expectation values of the field
squared and the energy-momentum tensor in this region are discussed in Sec.~%
\ref{sec:VEVex}. Sec. \ref{sec:Gen} generalizes the results for the exterior
region to a more general class of spherically-symmetric spacetimes inside a
spherical boundary. Section \ref{sec:Conc} contains a summary of the work.
In Appendix \ref{sec:AppBF} the expression for the Wightman function in
boundary-free dS spacetime is derived by making use the corresponding mode
sum. In Appendix \ref{sec:AppMink}, for the Wightman function, we explicitly
demonstrate the limit to the geometry of a spherical boundary in Minkowski
spacetime. Appendix \ref{app:C} sketches the corresponding calculation of
the Green's function for this problem.

\section{Wightman function inside a sphere}

\label{sec:WF}

We consider a quantum scalar field $\varphi (x)$ on a $(D+1)$-dimensional dS
spacetime background described in inflationary coordinates. The latter are
most appropriate for cosmological applications. The spatial part of the line
element we will write in terms of the hyperspherical coordinates $%
(r,\vartheta ,\phi )\equiv (r,\theta _{1},\theta _{2},\ldots \theta
_{n},\phi )$, $n=D-2$:
\begin{equation}
ds^{2}=dt^{2}-e^{2t/\alpha }(dr^{2}+r^{2}d\Omega _{D-1}^{2}),
\label{ds2deSit}
\end{equation}%
where $d\Omega _{D-1}^{2}$ is the line element on a $(D-1)$-dimensional
sphere with unit radius. In what follows, in addition to the synchronous
time coordinate, $t$, we will also use the conformal time, $\tau $, defined
as $\tau =-\alpha e^{-t/\alpha }$, $-\infty <\tau <0$. In terms of this
coordinate the line element takes the conformally flat form:%
\begin{equation}
ds^{2}=\alpha ^{2}\tau ^{-2}\left( d\tau ^{2}-dr^{2}-r^{2}d\Omega
_{D-1}^{2}\right) .  \label{ds2Dd}
\end{equation}%
Note that the parameter $\alpha $ is related to the cosmological constant $%
\Lambda $ through the expression $\alpha =D(D-1)/(2\Lambda )$.

The dynamics of a massive scalar field are governed by the equation \cite%
{Birr82}%
\begin{equation}
(\nabla _{l}\nabla ^{l}+m^{2}+\xi R)\varphi =0,  \label{fieldeq}
\end{equation}%
where $\nabla _{l}$ is the covariant derivative operator, $R=D(D+1)/\alpha
^{2}$ is the Ricci scalar for dS spacetime, and $\xi $ is the curvature
coupling parameter. The special values of this parameter $\xi =0$ and $\xi
=\xi _{D}\equiv (D-1)/4D$ correspond to minimally and to conformally coupled
fields. The importance of these two special cases comes from the fact that,
in the massless limit, the corresponding fields mimic the behavior of
gravitons and photons, respectively. Our main interest in this paper is the
study of the changes in the vacuum expectation values (VEVs) of the field
squared and the energy-momentum tensor induced by a spherical shell with
radius $a$ centered at the origin in dS spacetime. We assume that on the
sphere the field obeys Robin boundary condition%
\begin{equation}
(\tilde{A}+\tilde{B}\partial _{r})\varphi (x)=0,\quad r=a,  \label{spherebc}
\end{equation}%
with constant coefficients $\tilde{A}$ and $\tilde{B}$, in general,
different for the inner and outer regions. The results for Dirichlet and
Neumann boundary conditions are obtained as special cases. Robin boundary
conditions are an extension of the ones imposed on perfectly conducting
boundaries and may, in some geometries, be useful for depicting the finite
penetration of the field into the boundary with the ``skin-depth'' parameter
related to the Robin coefficient \cite{Most85}. These types of conditions
naturally arise for scalar and fermion bulk fields in braneworld models.

As the first step in the investigation of the VEVs we will evaluate the
Wightman function $W(x,x^{\prime })=\langle 0|\varphi (x)\varphi (x^{\prime
})|0\rangle $, where $|0\rangle $ stands for the vacuum state (for the
Wightman function in the geometry of spherical boundaries in the background
of a constant negative curvature space see Ref.~\cite{Saha08Neg}). In order
to do that we employ the mode sum formula
\begin{equation}
W(x,x^{\prime })=\sum_{\sigma }\varphi _{\sigma }(x)\varphi _{\sigma }^{\ast
}(x^{\prime }),  \label{WF}
\end{equation}%
with $\left\{ \varphi _{\sigma }(x),\varphi _{\sigma }^{\ast }(x)\right\} $
being a complete set of solutions to the classical field equation, specified
by a set of quantum numbers $\sigma $, satisfying the boundary condition (%
\ref{spherebc}). In accordance with the spherical symmetry of the problem
under consideration, the angular dependence of the mode functions is given
by the spherical harmonic of degree $l$ (see Ref.~\cite{Erdelyi}), $%
Y(m_{p};\vartheta ,\phi )$, where $m_{p}=(m_{0}\equiv l,m_{1},\ldots ,m_{n})$%
, $l=0,1,2,\ldots $, and $m_{1},m_{2},\ldots ,m_{n}$ are integers such that
\begin{eqnarray}
&& 0 \leqslant m_{n-1}\leqslant m_{n-2}\leqslant \cdots \leqslant
m_{1}\leqslant l,  \notag \\
&& -m_{n-1} \leqslant m_{n}\leqslant m_{n-1}.  \label{numbvalues}
\end{eqnarray}

Presenting the mode functions in the form $\varphi _{\sigma }(x)=T(\tau )%
\mathcal{R}(r)Y(m_{p};\vartheta ,\phi )$, from the field equation (\ref%
{fieldeq}) it follows that the time and the radial coordinate dependences
are given in terms of cylinder functions as:
\begin{eqnarray}
\mathcal{R}(r) &=&r^{1-D/2}\left[ b_{1}J_{\mu }(\lambda r)+b_{2}Y_{\mu
}(\lambda r)\right] ,  \notag \\
T(\tau ) &=&\eta ^{D/2}\sum_{j=1,2}c_{j}H_{\nu }^{(j)}(\lambda \eta ),\;\eta
=|\tau |,  \label{TR}
\end{eqnarray}%
where $J_{\mu }(z)$ and $Y_{\mu }(z)$ are the Bessel and Neumann functions
respectively, $H_{\nu }^{(j)}(z)$, $j=1,2$, are the Hankel functions (we use
the notations from Ref.~\cite{Abra72}). The orders of the cylinder functions
in Eq.~(\ref{TR}) are defined as:%
\begin{eqnarray}
\mu &=&l+D/2-1,  \notag \\
\nu &=&\left[ D^{2}/4-D(D+1)\xi -m^{2}\alpha ^{2}\right] ^{1/2}.
\label{munu}
\end{eqnarray}%
Note that $\nu $ is either real and nonnegative or purely imaginary. For a
conformally coupled massless field $\nu =1/2$ and the Hankel functions in
Eq.~(\ref{TR}) are expressed in terms of elementary functions.

Different choices of the coefficients $c_{j}$ in the expression for the
function $T(\tau )$ correspond to different choices of the vacuum state in
dS spacetime. The choice of the vacuum state is among the most important
steps in construction of a quantum field theory in a fixed classical
gravitational background. dS spacetime is a maximally symmetric space and it
is natural to choose a vacuum state having the same symmetry. In fact, there
exists a one-parameter family of maximally symmetric quantum states (see,
for instance, Ref.~\cite{Alle85} and references therein). Here we will
assume that the field is prepared in the dS-invariant Bunch-Davies vacuum
state \cite{Bunc78} for which $c_{2}=0$. Among the set of dS-invariant
quantum states the Bunch-Davies vacuum is the only one for which the
ultraviolet behavior of the two-point functions is the same as in Minkowski
spacetime.

First we consider the region inside the spherical shell. From the regularity
condition at the origin it follows that for this region $b_{2}=0$ and the
mode functions realizing the Bunch-Davies vacuum state are written in the
form%
\begin{equation}
\varphi _{\sigma }(x)=C_{\sigma }\frac{\eta ^{D/2}}{r^{D/2-1}}H_{\nu
}^{(1)}(\lambda \eta )J_{\mu }(\lambda r)Y(m_{p};\vartheta ,\phi ).
\label{eigfunc}
\end{equation}%
From the boundary condition (\ref{spherebc}), it follows that the
eigenvalues for $\lambda $ have to be solutions to the equation
\begin{equation}
AJ_{\mu }(\lambda a)+B\lambda aJ_{\mu }^{\prime }(\lambda a)=0,
\label{eigenmodes}
\end{equation}%
where the prime means the derivative with respect to the argument of the
function. In Eq.~(\ref{eigenmodes}) and in what follows we use the notations%
\begin{equation}
A=\tilde{A}+\left( 1-D/2\right) \tilde{B}/a,\quad B=\tilde{B}/a.  \label{AB}
\end{equation}%
For real $A$, $B$ and $\mu >-1$, all roots of Eq.~(\ref{eigenmodes}) are
simple and real, except the case $A/B<-\mu $ when there are two purely
imaginary zeros (see, e.g., Ref.~\cite{Watson}). We will denote by $%
z=\lambda _{\mu ,k},\,k=1,2,\ldots ,$ the zeros of the function $AJ_{\mu
}(z)+BzJ_{\mu }^{\prime }(z)$ in the right half-plane, assuming that they
are arranged in ascending order. So, for the eigenvalues of $\lambda $ one
has $\lambda =\lambda _{\mu ,k}/a$. Now we see that the set of quantum
numbers $\sigma $ is specified to $\sigma =(k,l,m_{1},\ldots ,m_{n})$. Note
that for Dirichlet boundary condition $A=1$, $B=0$, and for Neumann boundary
condition $A=(1-D/2)B$.

The coefficient $C_{\sigma }$ in Eq. (\ref{eigfunc}) is determined from the
orthonormalization condition \cite{Birr82}
\begin{equation}
-i\int d^{D}x\sqrt{|g|}g^{00}\varphi _{\sigma }(x)\overleftrightarrow{%
\partial }_{\tau }\varphi _{\sigma ^{\prime }}^{\ast }(x)=\delta _{\sigma
\sigma ^{\prime }},  \label{Normaliz}
\end{equation}%
where the integration goes over the region inside the sphere and $\delta
_{\sigma \sigma ^{\prime }}$ is the Kronecker delta. Substituting the
functions from Eq.~(\ref{eigfunc}), the integration over the angular
variables is performed by making use of the normalization integral for the
spherical harmonics:
\begin{equation}
\int \left\vert Y(m_{k};\vartheta ,\phi )\right\vert ^{2}d\Omega =N(m_{k}),
\label{harmint}
\end{equation}%
where the explicit form for $N(m_{k})$ can be found in Ref.~\cite{Erdelyi}.
By using also the relation (for $\nu $ purely real or imaginary)
\begin{equation}
H_{\nu }^{(1)}(\lambda \eta )H_{\nu ^{\ast }}^{(2)\prime }(\lambda \eta
)-H_{\nu ^{\ast }}^{(2)}(\lambda \eta )H_{\nu }^{(1)\prime }(\lambda \eta )=-%
\frac{4i}{\pi \lambda \eta }e^{-i(\nu -\nu ^{\ast })\pi /2},  \label{HankRel}
\end{equation}%
for the normalization coefficient we get%
\begin{equation}
C_{\sigma }^{2}=\frac{\pi \lambda T_{\mu }(\lambda a)e^{i(\nu -\nu ^{\ast
})\pi /2}}{2N(m_{k})a\alpha ^{D-1}},  \label{C2}
\end{equation}%
with the notation
\begin{equation}
T_{\mu }(z)=\frac{z}{(z^{2}-\mu ^{2})J_{\mu }^{2}(z)+z^{2}J_{\mu }^{^{\prime
}2}(z)}.  \label{Tmu}
\end{equation}

Substituting the eigenfunctions into the mode sum formula (\ref{WF}), for
the Wightman function one gets ($n=D-2$)%
\begin{eqnarray}
W(x,x^{\prime }) &=&\frac{\pi e^{i(\nu -\nu ^{\ast })\pi /2}}{2\alpha
^{D-1}nS_{D}a^{2}}\frac{\left( \eta \eta ^{\prime }\right) ^{D/2}}{%
(rr^{\prime })^{n/2}}\sum_{l=0}^{\infty }(2l+n)C_{l}^{n/2}(\cos \theta )
\notag \\
&&\times \sum_{k=1}^{\infty }zT_{\mu }(z)H_{\nu }^{(1)}(z\eta /a)H_{\nu
^{\ast }}^{(2)}(z\eta ^{\prime }/a)J_{\mu }(zr/a)J_{\mu }(zr^{\prime }/a)%
\big|_{z=\lambda _{\mu ,k}}.  \label{WF0}
\end{eqnarray}%
Here $C_{p}^{q}(x)$ is the Gegenbauer or ultraspherical polynomial of degree
$p$ and order $q$, $S_{D}=2\pi ^{D/2}/\Gamma (D/2)$ is the surface area of a
unit sphere in $D$-dimensional space, and $\theta $ is the angle between the
directions $(\vartheta ,\phi )$ and $(\vartheta ^{\prime },\phi ^{\prime })$%
. In deriving Eq. (\ref{WF0}), we have used the addition theorem \cite%
{Erdelyi} for the spherical harmonics:
\begin{equation}
\sum_{m_{p}}\frac{Y(m_{p};\vartheta ,\phi )}{N(m_{p})}Y^{\ast
}(m_{p};\vartheta ^{\prime },\phi ^{\prime })=\frac{2l+n}{nS_{D}}%
C_{l}^{n/2}(\cos \theta ),  \label{adtheorem}
\end{equation}%
where the sum is taken over the integer values $m_{p},\,p=1,2,\ldots ,n$, in
accordance with Eq.~(\ref{numbvalues}).

Before proceeding to the evaluation of the Wightman function, we comment
about the realizability of the Bunch-Davies vacuum state. It is well known
that in dS spacetime without boundaries the Bunch-Davies vacuum state is not
a physically realizable state for ${\mathrm{Re}}${\textrm{\thinspace }}$\nu
\geqslant D/2$. The corresponding Wightman function contains infrared
divergences arising from long-wavelength modes. As it has been shown in Ref.~%
\cite{Ford77}, these divergences lead to inconsistencies with Einstein
equations and they cannot arise through dynamical evolution from a state
which is initially free of such divergences. In the presence of boundaries,
the boundary condition imposed on the quantized field may exclude these
modes and the Bunch-Davies vacuum becomes a realizable state. An example is
provided by a spherical boundary described above. In the region inside the
spheres and for boundary conditions with $\tilde{A}\neq 0$, there is a
minimum value for $\lambda $, $\lambda \geqslant \lambda _{n/2,1}/a$, and
the two-point function (\ref{WF0}) contains no infrared divergences.

The eigenvalues $\lambda _{\mu ,k}$ are given implicitly and the Wightman
function in the form (\ref{WF0}) is not convenient for the further
evaluation of the VEVs. In order to sum over these eigenvalues we use the
summation formula \cite{Saha08b}
\begin{eqnarray}
2\sum_{k=1}^{\infty }T_{\mu }(\lambda _{\mu ,k})f(\lambda _{\mu ,k})
&=&\int_{0}^{\infty }f(x)dx+\frac{\pi }{2}\mathrm{Res}_{z=0}f(z)\frac{\bar{Y}%
_{\mu }(z)}{\bar{J}_{\mu }(z)}{}  \notag \\
&&\quad\mbox{}-\frac{1}{\pi }\int_{0}^{\infty }\ dx\frac{\bar{K}_{\mu }(x)}{%
\bar{I}_{\mu }(x)}\left[ e^{-\mu \pi i}f(xe^{\pi i/2})+e^{\mu \pi
i}f(xe^{-\pi i/2})\right] ,  \label{sumJ1anal}
\end{eqnarray}%
where $f(z)$ is an analytic function on the right half-plane and, for a
given function $F(z)$, the barred notation is defined as:
\begin{equation}
\bar{F}(z)\equiv AF(z)+BzF^{\prime }(z).  \label{barnot}
\end{equation}%
Formula (\ref{sumJ1anal}) can be generalized for the case of the existence
of purely imaginary zeros of the function $\bar{J}_{\nu }(z)$ by adding the
corresponding residue term and taking the principal value of the integral on
the right (see Ref.~\cite{Saha08b}). In what follows we assume values of $%
A/B $ for which all roots $\lambda _{\mu ,k}$ are real.

As a function $f(z)$ in Eq.~(\ref{sumJ1anal}) we take
\begin{equation}
f(z)=zH_{\nu }^{(1)}(z\eta /a)H_{\nu ^{\ast }}^{(2)}(z\eta ^{\prime
}/a)J_{\mu }(zr/a)J_{\mu }(zr^{\prime }/a).  \label{fz}
\end{equation}%
By making use of the properties of the cylinder functions, after the
application of Eq.~(\ref{sumJ1anal}), the Wightman function is presented in
the decomposed form:
\begin{equation}
W(x,x^{\prime })=W_{\text{dS}}(x,x^{\prime })+W_{\text{b}}(x,x^{\prime }).
\label{WF1}
\end{equation}%
Here, the first term in the right-hand side corresponds to the first
integral on the right of Eq. (\ref{sumJ1anal}),%
\begin{eqnarray}
W_{\text{dS}}(x,x^{\prime }) &=&\frac{\pi e^{i(\nu -\nu ^{\ast })\pi /2}}{%
4\alpha ^{D-1}nS_{D}}\frac{\left( \eta \eta ^{\prime }\right) ^{D/2}}{%
(rr^{\prime })^{n/2}}\sum_{l=0}^{\infty }(2l+n)C_{l}^{n/2}(\cos \theta )
\notag \\
&&\times \int_{0}^{\infty }d\lambda \,\lambda H_{\nu }^{(1)}(\lambda \eta
)H_{\nu ^{\ast }}^{(2)}(\lambda \eta ^{\prime })J_{\mu }(\lambda r)J_{\mu
}(\lambda r^{\prime }),  \label{WdS}
\end{eqnarray}%
and
\begin{eqnarray}
W_{\text{b}}(x,x^{\prime }) &=&-\frac{\alpha ^{1-D}}{\pi nS_{D}}\frac{\left(
\eta \eta ^{\prime }\right) ^{D/2}}{(rr^{\prime })^{n/2}}\sum_{l=0}^{\infty
}(2l+n)C_{l}^{n/2}(\cos \theta )\int_{0}^{\infty }\ dz\,\,z  \notag \\
&&\times \frac{\bar{K}_{\mu }(za)}{\bar{I}_{\mu }(za)}I_{\mu }(zr)I_{\mu
}(zr^{\prime })\left[ I_{\nu }(z\eta ^{\prime })K_{\nu }(z\eta )+I_{-\nu
}(z\eta )K_{\nu }(z\eta ^{\prime })\right] .  \label{Wb}
\end{eqnarray}%
The term (\ref{WdS}) does not depend on the radius of the sphere whereas the
second term on the right-hand side vanishes in the limit $a\rightarrow
\infty $. From here it follows that $W_{\text{dS}}(x,x^{\prime })$ is the
Wightman function for a scalar field in boundary-free dS spacetime. In
Appendix \ref{sec:AppBF} we show this by direct evaluation. The second term
in the right-hand side of Eq. (\ref{WF1}) is induced by the presence of the
spherical shell. In the limit $\alpha \rightarrow \infty $ with fixed $t$,
the line element (\ref{ds2deSit}) goes to the Minkowskian line element in
spherical coordinates. In Appendix \ref{sec:AppMink} it is shown that in
this limit, from Eq.~(\ref{Wb}), the Wightman function is obtained for a
scalar field inside a spherical boundary in Minkowski spacetime. In Appendix
C we derive the corresponding causal Green's function.

In a similar way we can evaluate the Wightman function in a general state
described by the modes (\ref{TR}) without the specification of the
coefficients in the linear combination of the Hankel functions. After the
normalization of the corresponding mode functions, we obtain a set of
quantum states determined by the ratio $c_{2}/c_{1}$. In general, the latter
may depend on $l$ and $\lambda $. After the application of the generalized
Abel-Plana formula, the boundary-induced part in the corresponding Wigthman
function is presented in a form similar to Eq.~(\ref{Wb}) where now, instead
of the combination of functions in the square brackets, a more general
bilinear combination of the modified Bessel functions appears.

\section{VEV of the field squared inside a sphere}

\label{sec:phi2}

Given the Wightman function, we can proceed to the evaluation of the VEV of
the field squared. This VEV\ is among the most important quantities in
discussing the phase transitions in the early universe and the generation of
the cosmic structures from inflation. The VEV of the field squared is
obtained taking the coincidence limit of the arguments. In this limit the
Wightman function is divergent and some renormalization procedure is needed.
The important point here is that for points away from the sphere the
divergences are the same as those for dS spacetime without boundaries. As we
have already extracted the part $W_{\text{dS}}(x,x^{\prime })$, the
renormalization is reduced to the renormalization of the VEV in the
boundary-free dS spacetime, which is already done in the literature. In this
way, the renormalized VEV of the field squared inside a sphere is presented
in the decomposed form%
\begin{equation}
\langle \varphi ^{2}\rangle =\langle \varphi ^{2}\rangle _{\text{dS}%
}+\langle \varphi ^{2}\rangle _{\text{b}},  \label{phi2dec}
\end{equation}%
where $\langle \varphi ^{2}\rangle _{\text{dS}}$ is the VEV in the
boundary-free dS spacetime and the term $\langle \varphi ^{2}\rangle _{\text{%
b}}$ is induced by the sphere. Due to the maximal symmetry of the
Bunch-Davies vacuum state the VEV $\langle \varphi ^{2}\rangle _{\text{dS}}$
does not depend on the spacetime point.

The boundary-induced part is directly obtained from the corresponding part
in the Wightman function, given by Eq. (\ref{Wb}). By taking into account
that
\begin{equation}
C_{l}^{n/2}(1)=\frac{\Gamma (l+n)}{\Gamma (n)l!},  \label{Cl1}
\end{equation}%
we get%
\begin{equation}
\langle \varphi ^{2}\rangle _{\text{b}}=-\frac{\eta ^{D}\alpha ^{1-D}}{\pi
S_{D}r^{D-2}}\sum_{l=0}^{\infty }D_{l}\int_{0}^{\infty }dx\,x\frac{\bar{K}%
_{\mu }(ax)}{\bar{I}_{\mu }(ax)}I_{\mu }^{2}(rx)K_{\nu }(x\eta )\left[
I_{\nu }(x\eta )+I_{-\nu }(x\eta )\right] .  \label{phi2b}
\end{equation}%
In Eq. (\ref{phi2b}),
\begin{equation}
D_{l}=(2l+D-2)\frac{\Gamma (l+D-2)}{\Gamma (D-1)\,l!}  \label{Dlang}
\end{equation}%
is the degeneracy of the angular mode with given $l$. The integral
representation (\ref{phi2b}) is valid for $\mathrm{Re\,}\nu <1$. For large
values of $x$, the integrand in Eq. (\ref{phi2b}) behaves as $e^{-(a-r)x}$.
The presence of the spherical boundary breaks the dS-invariance and the
mean-squared fluctuation of the field depends on time. As is seen from Eq.~(%
\ref{phi2b}), this dependence appears through the ratios $a/\eta $ and $%
r/\eta $. The latter property is a consequence of the maximal symmetry of
the Bunch-Davies vacuum in the absence of the sphere. Note that $a/\eta $
and $r/\eta $ are the proper radius of the sphere and the proper distance
from the sphere center measured in units of the dS curvature scale $\alpha $%
. The influence of the gravitational field on boundary-induced quantum
effects appears through the function $K_{\nu }(z)\left[ I_{\nu }(z)+I_{-\nu
}(z)\right] $. The latter is a monotonically decreasing function of $z$ for $%
\nu ^{2}\geqslant 0$ and exhibits an oscillatory behavior for $\nu ^{2}<0$
and $z\lesssim |\nu |$. As it will be seen below, this feature leads to
interesting physical consequences.

For a conformally coupled massless scalar field, $\xi =\xi _{D}$, $m=0$, one
has $\nu =1/2$ and%
\begin{equation}
K_{1/2}(x)\left[ I_{1/2}(x)+I_{-1/2}(x)\right] =1/x.  \label{KIconf}
\end{equation}%
In this case, from Eq.~(\ref{phi2b}), we find%
\begin{equation}
\langle \varphi ^{2}\rangle _{\text{b}}=(\eta /\alpha )^{D-1}\langle \varphi
^{2}\rangle _{\text{M}},  \label{phi2conf}
\end{equation}%
where%
\begin{equation}
\langle \varphi ^{2}\rangle _{\text{M}}=-\frac{r^{2-D}}{\pi S_{D}}%
\sum_{l=0}^{\infty }D_{l}\int_{0}^{\infty }dx\,\frac{\bar{K}_{\mu }(ax)}{%
\bar{I}_{\mu }(ax)}I_{\mu }^{2}(rx),  \label{phi2M}
\end{equation}%
is the corresponding VEV inside a spherical boundary in Minkowski spacetime
\cite{Saha01}. Of course, this result could be obtained directly by using
the conformal relation between the problems in dS and Minkowski spacetimes.

Now let us consider the asymptotics of the boundary-induced VEV. Near the
center of the sphere the dominant contribution comes from the mode with the
lowest orbital momentum $l=0$ and in the leading order we find%
\begin{equation}
\langle \varphi ^{2}\rangle _{\text{b}}\approx -\frac{(2\alpha )^{1-D}}{\pi
^{D/2+1}\Gamma (D/2)}\int_{0}^{\infty }dx\,x^{D-1}\frac{\bar{K}%
_{D/2-1}(ax/\eta )}{\bar{I}_{D/2-1}(ax/\eta )}K_{\nu }(x)\left[ I_{\nu
}(x)+I_{-\nu }(x)\right] .  \label{phi2center}
\end{equation}%
The behavior of the field squared for large values of the sphere proper
radius, $a/\eta \gg 1$, assuming that $r/\eta $ is fixed, is obtained from
Eq.~(\ref{phi2center}) expanding the function $K_{\nu }(x)\left[ I_{\nu
}(x)+I_{-\nu }(x)\right] $ for small values of the argument. For $x\ll 1$ in
the leading order one has%
\begin{equation}
K_{\nu }(x)\left[ I_{\nu }(x)+I_{-\nu }(x)\right] \approx \sigma _{\nu }{%
\mathrm{Re}}\left[ \frac{2^{2\nu -1}\Gamma (\nu )}{\Gamma (1-\nu )x^{2\nu }}%
\right] ,  \label{KIas}
\end{equation}%
where $\sigma _{\nu }=1$ for positive $\nu $ and $\sigma _{\nu }=2$ for
imaginary $\nu $. The case $\nu =0$ should be considered separately. The
behavior of the VEV is qualitatively different for these two cases. For
positive $\nu $, one gets%
\begin{equation}
\langle \varphi ^{2}\rangle _{\text{b}}\approx -\frac{(2\alpha )^{1-D}}{\pi
^{D/2+1}}\frac{(\eta /a)^{D-2\nu }}{\Gamma (D/2)}b(\nu ),  \label{phi2Larga}
\end{equation}%
with the notation%
\begin{equation}
b(\nu )=\frac{2^{2\nu -1}\Gamma (\nu )}{\Gamma (1-\nu )}\int_{0}^{\infty
}dx\,x^{D-1-2\nu }\frac{\bar{K}_{D/2-1}(x)}{\bar{I}_{D/2-1}(x)}.  \label{bnu}
\end{equation}%
In this case the boundary-induced VEV monotonically decreases with
increasing proper radius of the sphere. For imaginary values of $\nu $, the
leading term is in the form%
\begin{equation}
\langle \varphi ^{2}\rangle _{\text{b}}\approx -\frac{2(2\alpha )^{1-D}}{\pi
^{D/2+1}}\frac{(\eta /a)^{D}}{\Gamma (D/2)}c(\nu )\cos [2|\nu |\ln (a/\eta
)+\phi (\nu )],  \label{phi2LargaIm}
\end{equation}%
where $c(\nu )$ and $\phi (\nu )$ are defined by the relation $b(\nu )=$ $%
c(\nu )e^{i\phi (\nu )}$. In this case the VEV exhibits a damped oscillatory
behavior.

The VEV of the field squared diverges on the boundary. Surface divergences
are well known in quantum field theory wtih boundaries and they have been
investigated for various geometries. Near the sphere the dominant
contribution in Eq.~(\ref{phi2b}) comes from large values of $x$ and $l$. By
taking into account that for large $x$ and for fixed $\nu $ one has $K_{\nu
}(x)\left[ I_{\nu }(x)+I_{-\nu }(x)\right] \approx 1/x$, we conclude that
near the boundary, to the leading order, one has $\langle \varphi
^{2}\rangle _{\text{b}}\approx (\eta /\alpha )^{D-1}\langle \varphi
^{2}\rangle _{\text{M}}$. By using the asymptotic expression for $\langle
\varphi ^{2}\rangle _{\text{M}}$, we get%
\begin{equation}
\langle \varphi ^{2}\rangle _{\text{b}}\approx -\frac{(\eta /\alpha
)^{D-1}\Gamma ((D-1)/2)}{(4\pi )^{(D+1)/2}(a-r)^{D-1}}\kappa _{B},
\label{phi2near}
\end{equation}%
where%
\begin{equation}
\kappa _{B}=2\delta _{0B}-1.  \label{kappaB}
\end{equation}%
In deriving Eq.~(\ref{phi2near}) we have assumed that $(a-r)\ll \eta
/(m\alpha )$ which corresponds to small proper distances from the sphere
compared with the Compton wavelength of the scalar particle. As the
boundary-free part in the VEV is constant everywhere, we conclude that near
the sphere the total VEV is dominated by the boundary-induced part. The
factor $\kappa _{B}$ in Eq.~(\ref{phi2near}) indicates that the sign of the
surface divergence is reversed for Dirichlet as opposed to any other Robin
boundary condition.

\section{Vacuum energy-momentum tensor inside a sphere}

\label{sec:EMT}

Another important quantity characterizing the properties of the quantum
vacuum is the VEV of the energy-momentum tensor. In addition to describing
the physical structure of the quantum field at a given point, the
energy-momentum tensor acts as the source of gravity in the Einstein
equations. It therefore plays an important role in modelling a
self-consistent dynamics involving the gravitational field. (For the
renormalization of the Einstein equations in the presence of energy-momentum
tensor divergences, see Ref.~\cite{estrada}; the demonstration that the
finite and divergent parts of the Casimir energy obey the equivalence
principle appears in Refs.~\cite{fall1,fall2}.) Similar to the field
squared, the VEV of the energy-momentum tensor is decomposed as%
\begin{equation}
\langle T_{ik}\rangle =\langle T_{ik}\rangle _{\text{dS}}+\langle
T_{ik}\rangle _{\text{b}},  \label{TikDec}
\end{equation}%
where the boundary-free part is presented in the form $\langle T_{ik}\rangle
_{\text{dS}}=$const$\cdot g_{ik}$. The latter property is a direct
consequence of the maximal symmetry of the Bunch-Davies vacuum state. In
particular, for $D=3$ dS spacetime the renormalized boundary-free part is
given by the expression \cite{Bunc78,Cand75} (see also Ref.~\cite{Birr82})
\begin{eqnarray}
\langle T_{ik}\rangle _{\text{dS}} &=&\frac{g_{ik}}{32\pi ^{2}\alpha ^{4}}%
\{m^{2}\alpha ^{2}\left( m^{2}\alpha ^{2}/2+6\xi -1\right) \left[ \psi
\left( 3/2+\nu \right) +\psi \left( 3/2-\nu \right) -\ln \left( m^{2}\alpha
^{2}\right) \right]  \notag \\
&&-\left( 6\xi -1\right) ^{2}+1/30+(2/3-6\xi )m^{2}\alpha ^{2}\},
\label{Tik0ren}
\end{eqnarray}%
where $\psi (x)$ is the logarithmic derivative of the gamma-function. For $%
m\alpha \gg 1$, to the leading order, $\langle T_{ik}\rangle _{\text{dS}%
}\approx Cg_{ik}/(384\pi ^{2}m^{2}\alpha ^{6})$, where the coefficient $C$
depends on the curvature coupling parameter only. For minimally and
conformally coupled fields we have $C=7$ and $C=-1/5$ respectively.

For the evaluation of the boundary-induced part in the VEV of the
energy-momentum tensor, we use the formula%
\begin{equation}
\langle T_{ik}\rangle _{\text{b}}=\lim_{x^{\prime }\rightarrow x}\partial
_{i}\partial _{k}^{\prime }W_{\text{b}}(x,x^{\prime })+\left[ \left( \xi
-1/4\right) g_{ik}\nabla _{l}\nabla ^{l}-\xi \nabla _{i}\nabla _{k}-\xi
R_{ik}\right] \langle \varphi ^{2}\rangle _{\text{b}},  \label{emtvev1}
\end{equation}%
where $R_{ik}=Dg_{ik}/\alpha ^{2}$ is the Ricci tensor for dS spacetime. By
taking into account Eqs. (\ref{Wb}) and (\ref{phi2b}), after long but
straightforward calculations, the VEVs for the diagonal components are
presented in the form (no summation over $i$)%
\begin{equation}
\langle T_{i}^{i}\rangle _{\text{b}}=-\frac{r^{2-D}\eta ^{2}}{2\pi
S_{D}\alpha ^{D+1}}\sum_{l=0}^{\infty }D_{l}\int_{0}^{\infty }dx\,x^{3-D}%
\frac{\bar{K}_{\mu }(ax)}{\bar{I}_{\mu }(ax)}\left\{ G_{i}^{i}[I_{\mu
}(rx)]F_{\nu }(x\eta )+2I_{\mu }^{2}(rx)F_{i}^{i}(x\eta )\right\} ,
\label{Tiib}
\end{equation}%
where we have introduced the notation%
\begin{equation}
F_{\nu }(z)=z^{D}K_{\nu }(z)\left[ I_{\nu }(z)+I_{-\nu }(z)\right] .
\label{Fnu}
\end{equation}%
Other functions in Eq.~(\ref{Tiib}) are defined in accordance with (no
summation over $k$)%
\begin{eqnarray}
F_{0}^{0}(y) &=&\left[ \frac{1}{4}\partial _{y}^{2}-D\frac{\xi +\xi _{D}}{y}%
\partial _{y}-1+\frac{D^{2}\xi +m^{2}\alpha ^{2}}{y^{2}}\right] F_{\nu }(y),
\notag \\
F_{k}^{k}(y) &=&\left[ \left( \xi -\frac{1}{4}\right) \partial _{y}^{2}+%
\left[ \xi (2-D)+\frac{D-1}{4}\right] \frac{1}{y}\partial _{y}-\frac{\xi D}{%
y^{2}}\right] F_{\nu }(y),  \label{Fii}
\end{eqnarray}%
for $k=1,2,\ldots ,D$, and
\begin{eqnarray}
G_{0}^{0}\left[ f(y)\right] &=&\left( 1-4\xi \right) \left[ f^{^{\prime
}2}(y)-\frac{D-2}{y}f(y)f^{\prime }(y)+\left( 1+\frac{\mu ^{2}}{y^{2}}%
\right) f^{2}(y)\right] ,  \notag \\
G_{1}^{1}\left[ f(y)\right] &=&f^{\prime 2}(y)+\frac{\xi _{1}}{y}%
f(y)f^{\prime }(y)-\left[ 1+\frac{(D/2-1)\xi _{1}+\mu ^{2}}{y^{2}}\right]
f^{2}(y),  \label{Gii} \\
G_{k}^{k}\left[ f(y)\right] &=&\left( 4\xi -1\right) f^{\prime 2}(y)-\frac{%
\xi _{1}}{y}f(y)f^{\prime }(y)+\left[ 4\xi -1+\frac{(D/2-1)\xi _{1}+\mu
^{2}(1+\xi _{1})}{(D-1)y^{2}}\right] f^{2}(y),  \notag
\end{eqnarray}%
for $k=2,\ldots ,D$, where%
\begin{equation}
\xi _{1}=4(D-1)\xi -D+2.  \label{ksi1}
\end{equation}

In addition, the boundary-induced VEV has also nonzero off-diagonal component%
\begin{eqnarray}
\langle T_{0}^{1}\rangle _{\text{b}} &=&\frac{\eta \alpha ^{-D-1}}{\pi
S_{D}r^{D-1}}\sum_{l=0}^{\infty }D_{l}\int_{0}^{\infty }dx\,x^{1-D}\frac{%
\bar{K}_{\mu }(ax)}{\bar{I}_{\mu }(ax)}I_{\mu }(rx)  \notag \\
&&\times \left[ 2rxI_{\mu }^{\prime }(rx)-(D-2)I_{\mu }(rx)\right] \left[
\left( \xi -1/4\right) \eta \partial _{\eta }+\xi \right] F_{\nu }(x\eta ),
\label{T01b}
\end{eqnarray}%
which corresponds to the energy flux along the radial direction (see below).
Note that in the formulas above, the components of the energy-momentum
tensor are written in the coordinates $(\tau ,r,\vartheta ,\phi )$. The
components in the system $(t,r,\vartheta ,\phi )$ with the synchronous time
coordinate will be denoted by $\langle T_{\text{(s)}i}^{k}\rangle _{\text{b}%
} $. We have the relations (no summation over $i$)%
\begin{equation}
\langle T_{\text{(s)}i}^{i}\rangle _{\text{b}}=\langle T_{i}^{i}\rangle _{%
\text{b}},\;\langle T_{\text{(s)}0}^{1}\rangle _{\text{b}}=(\eta /\alpha
)\langle T_{0}^{1}\rangle _{\text{b}}.  \label{ConfSyncrel}
\end{equation}

As an additional check of the expressions for the VEV of the energy-momentum
tensor, it can be seen that the boundary-induced parts fulfill the trace
relation%
\begin{equation}
\langle T_{l}^{l}\rangle _{\text{b}}=\left[ D(\xi -\xi _{D})\nabla
_{l}\nabla ^{l}+m^{2}\right] \langle \varphi ^{2}\rangle _{\text{b}}.
\label{TrRel}
\end{equation}%
In particular, the boundary-induced part in the VEV of the energy-momentum
tensor is traceless for a conformally coupled massless scalar field. The
trace anomaly is contained in the boundary-free part only. In addition, the
VEVs obey the covariant conservation equation $\nabla _{k}\langle
T_{i}^{k}\rangle _{\text{b}}=0$. For the geometry of interest, this equation
is reduced to two equations:
\begin{eqnarray}
\left( \partial _{\eta }-\frac{D}{\eta }\right) \langle T_{0}^{0}\rangle _{%
\text{b}}-\partial _{r}\langle T_{0}^{1}\rangle _{\text{b}}+\frac{1}{\eta }%
\left[ \langle T_{1}^{1}\rangle _{\text{b}}+(D-1)\langle T_{2}^{2}\rangle _{%
\text{b}}\right] &=&0,  \notag \\
\left( \partial _{\eta }-\frac{D+1}{\eta }\right) \langle T_{1}^{0}\rangle _{%
\text{b}}-\left( \partial _{r}+\frac{D-1}{r}\right) \langle T_{1}^{1}\rangle
_{\text{b}}+\frac{D-1}{r}\langle T_{2}^{2}\rangle _{\text{b}} &=&0.
\label{ConsEq}
\end{eqnarray}

For a conformally coupled massless field one has $\nu =1/2$ and $F_{\nu
}(z)=z^{D-1}$. With this it is easily seen that the off-diagonal component
vanishes and $F_{0}^{0}(y)=-y^{D-1}$, $F_{k}^{k}(y)=0$ for $k=1,2,\ldots ,D$%
. As a result we find (no summation over $i$)%
\begin{equation}
\langle T_{i}^{i}\rangle _{\text{b}}=\left( \eta /\alpha \right)
^{D+1}\langle T_{i}^{i}\rangle _{\text{M}},  \label{Tiiconf}
\end{equation}%
where $\langle T_{i}^{i}\rangle _{\text{M}}$ is the corresponding VEV for a
sphere in Minkowski spacetime \cite{Saha01}. Again, the result (\ref{Tiiconf}%
) directly follows from the conformal relation between the problems in dS
and Minkowski spacetimes. The electromagnetic field is conformally invariant
in 4-dimensional spacetimes. Hence, the electromagnetic Casimir densities
for a conducting spherical shell on background of $D=3$ dS spacetime are
obtained from those in Minkowski spacetime by using Eq. (\ref{Tiiconf}).

Let us denote by $E_{V}^{\text{(b)}}$ the boundary-induced part of the
vacuum energy in the spatial volume $V$ with a boundary $\partial V$:%
\begin{equation}
E_{V}^{\text{(b)}}=\int_{V}d^{D}x\sqrt{\gamma }\langle T_{\text{(s)}%
0}^{0}\rangle _{\text{b}},  \label{EbV}
\end{equation}%
where $\gamma $ is the determinant of the spatial metric tensor $\gamma
_{\beta \delta }$ with $g_{ik}=(1,-\gamma _{\beta \delta })$ and with the
Greek indices running over $1,2,\ldots ,D$. Now, from the covariant
conservation equation $\nabla _{k}\langle T_{\text{(s)}i}^{k}\rangle _{\text{%
b}}=0$ with $i=0$, it follows that%
\begin{equation}
\partial _{t}E_{V}^{\text{(b)}}=-\int_{\partial V}d^{D-1}x\,\sqrt{h}n_{\beta
}\langle T_{\text{(s)}0}^{\beta }\rangle _{\text{b}}+\frac{1}{\alpha }%
\int_{V}d^{D}x\sqrt{\gamma }\langle T_{\text{(s)}\beta }^{\beta }\rangle _{%
\text{b}},  \label{EnCons}
\end{equation}%
where $n_{\beta }$, $\gamma ^{\beta \delta }n_{\beta }n_{\delta }=1$, is the
external normal to the boundary $\partial V$ and $h$ is the determinant of
the induced metric $h_{\beta \delta }=\gamma _{\beta \delta }-n_{\beta
}n_{\gamma }$. The first term in the right-hand side of Eq. (\ref{EnCons})
describes the energy flux through the boundary $\partial V$. In particular,
for a spherical boundary with radius $r_{0}<a$ and with the center at $r=0$
one has $n_{\beta }=(\alpha /\eta )\delta _{\beta }^{1}$ and the flux term
is given by $-(\alpha r_{0}/\eta )^{D-1}S_{D}(\alpha /\eta )\langle T_{\text{%
(s)}0}^{1}\rangle _{\text{b}}$. Now, by taking into account that the proper
surface area of the sphere with radius $r_{0}$ is given by $(\alpha
r_{0}/\eta )^{D-1}S_{D}$, we conclude that the quantity $\langle
T_{0}^{1}\rangle _{\text{b}}=(\alpha /\eta )\langle T_{\text{(s)}%
0}^{1}\rangle _{\text{b}}$, given by Eq. (\ref{T01b}), is the energy flux
density per unit proper surface area.

By using the expression (\ref{Tiib}) for the energy density, for the
boundary-induced vacuum energy inside the sphere with radius $r_{0}$ one gets%
\begin{eqnarray}
E_{r\leqslant r_{0}}^{\text{(b)}} &=&-\frac{(r_{0}/\eta )^{D-2}}{2\pi \alpha
}\sum_{l=0}^{\infty }D_{l}\int_{0}^{\infty }dx\,x^{1-D}\frac{\bar{K}_{\mu
}(ax/r_{0})}{\bar{I}_{\mu }(ax/r_{0})}  \notag \\
&&\big\{\left( 1-4\xi \right) F_{\nu }(x\eta /r_{0})I_{\mu }(x)\left[
x\,I_{\mu }^{\prime }(x)-\frac{D-2}{2}I_{\mu }(x)\right]  \notag \\
&&-F_{0}^{0}(x\eta /r_{0})\left[ x^{2}I_{\mu }^{\prime 2}(x)-(x^{2}+\mu
^{2})I_{\mu }^{2}(x)\right] \big\}.  \label{Etot}
\end{eqnarray}%
In deriving Eq. (\ref{Etot}) we have used the formula for the integral
involving the square of the modified Bessel function \cite{Prud86}. Related
to the surface divergences in the vacuum energy density (see below), the
integrated energy $E_{r\leqslant r_{0}}^{\text{(b)}}$ diverges in the limit $%
r_{0}\rightarrow a$. In order to obtain a finite result for the vacuum
energy inside the sphere, an additional renormalization procedure is needed.
Our main interest in the present paper are local characteristics of the
vacuum away from the boundary. As we have noted before, the renormalization
of the latter is reduced to that for the boundary-free dS geometry.

The general expressions (\ref{Tiib}) and (\ref{T01b}) are simplified at the
sphere's center and near the sphere. Near the center, the dominant
contribution to the boundary-induced VEVs comes from the modes with $l=0$
and $l=1$. To the leading order, for the diagonal components we get (no
summation over $i$)%
\begin{eqnarray}
\langle T_{i}^{i}\rangle _{\text{b}} &\approx &-\frac{2^{-D}\alpha ^{-D-1}}{%
\pi ^{D/2+1}\Gamma \left( D/2\right) }\int_{0}^{\infty }dx\,x\left\{ \frac{%
\bar{K}_{D/2}(ax/\eta )}{\bar{I}_{D/2}(ax/\eta )}G_{(1)}^{i}F_{\nu
}(x)\right.  \notag \\
&&\left. +\frac{\bar{K}_{D/2-1}(ax/\eta )}{\bar{I}_{D/2-1}(ax/\eta )}\left[
G_{(0)}^{i}F_{\nu }(x)+2F_{i}^{i}(x)\right] \right\} ,  \label{TiibCentr}
\end{eqnarray}%
where we have defined
\begin{eqnarray}
G_{(0)}^{0} &=&1-4\xi ,\;G_{(0)}^{k}=\frac{4(D-1)\xi }{D}-1,  \notag \\
G_{(1)}^{0} &=&D\frac{1-4\xi }{2},\;G_{(1)}^{k}=4(D-1)\xi -D+2,  \label{Gil}
\end{eqnarray}%
for $k=1,2,\ldots ,D$. The first and second terms in the figure braces of
Eq.~(\ref{TiibCentr}) come from the modes $l=1$ and $l=0$, respectively. As
we could expect, the stresses are isotropic at the center. For the
off-diagonal component of the VEV one has the following leading term:%
\begin{eqnarray}
\langle T_{0}^{1}\rangle _{\text{b}} &\approx &\frac{2^{1-D}\alpha
^{-D-1}\eta r}{\pi ^{D/2+1}\Gamma \left( D/2+1\right) }\int_{0}^{\infty
}dx\,x\left[ \frac{\bar{K}_{D/2-1}(ax)}{\bar{I}_{D/2-1}(ax)}+\frac{\bar{K}%
_{D/2}(ax)}{\bar{I}_{D/2}(ax)}\right]  \notag \\
&&\times \left[ \left( \xi -1/4\right) \eta \partial _{\eta }+\xi \right]
F_{\nu }(x\eta ).  \label{T01bCentr}
\end{eqnarray}%
The first and second terms in the square brackets are the contributions of
the modes $l=0$ and $l=1$. The off-diagonal component vanishes at the
center. The behavior of the energy-momentum tensor components for large
values of the sphere proper radius, $a/\eta \gg 1$, assuming fixed values of
$r/\eta $, is obtained from Eqs.~(\ref{TiibCentr}) and (\ref{T01bCentr}) by
expanding the function $F_{\nu }(z)$ for small values of the argument. By
using Eq.~(\ref{KIas}), similar to the case of the field squared, it can be
seen that for positive values of $\nu $ the VEVs decay monotonically like $%
(a/\eta )^{2\nu -D-2}$. For imaginary $\nu $, the behavior of the VEVs is
damped oscillatory with the amplitude decaying as $(a/\eta )^{-D-2}$.

For points near the surface of the sphere, the dominant contribution to the
boundary-induced VEVs comes from large values of $l$ and $x$. By using the
uniform asymptotic expansions for the modified Bessel functions, it can be
seen that the leading terms in the diagonal components for a scalar field
with non-conformal coupling ($\xi \neq \xi _{D}$) are related to the
corresponding terms for a spherical boundary in Minkowski spacetime by (no
summation over $i$) $\langle T_{i}^{i}\rangle _{\text{b}}\approx (\eta
/\alpha )^{D+1}\langle T_{i}^{i}\rangle _{\text{M}}$. These leading terms
are given by the expressions (no summation over $k$)%
\begin{eqnarray}
\langle T_{k}^{k}\rangle _{\text{b}} &\approx &\frac{D\Gamma ((D+1)/2)(\xi
-\xi _{D})\kappa _{B}}{2^{D}\pi ^{(D+1)/2}[\alpha (a-r)/\eta ]^{D+1}},
\notag \\
\langle T_{1}^{1}\rangle _{\text{b}} &\approx &\frac{(D-1)\Gamma
((D+1)/2)(\xi -\xi _{D})\kappa _{B}}{2^{D}\pi ^{(D+1)/2}a(a-r)^{D}(\alpha
/\eta )^{D+1}},  \label{TiiNear}
\end{eqnarray}%
for $k=0,2,\ldots ,D$, and $\kappa _{B}$ is defined in Eq. (\ref{kappaB}).
For the off-diagonal component we have%
\begin{equation}
\langle T_{0}^{1}\rangle _{\text{b}}\approx \frac{D\Gamma (D/2)( \xi -\xi
_{D}) \kappa _{B} }{2^{D}\pi ^{D/2+1}\alpha \lbrack \alpha (a-r)/\eta ]^{D}}.
\label{T10bNear}
\end{equation}%
The leading terms do not depend on the mass and have opposite signs for
Dirichlet and non-Dirichlet boundary conditions. In particular, for a
minimally coupled Dirichlet scalar field the energy density near the sphere
is negative and the energy flux is directed away from the boundary.

\begin{figure}[tbph]
\begin{center}
\begin{tabular}{cc}
\epsfig{figure=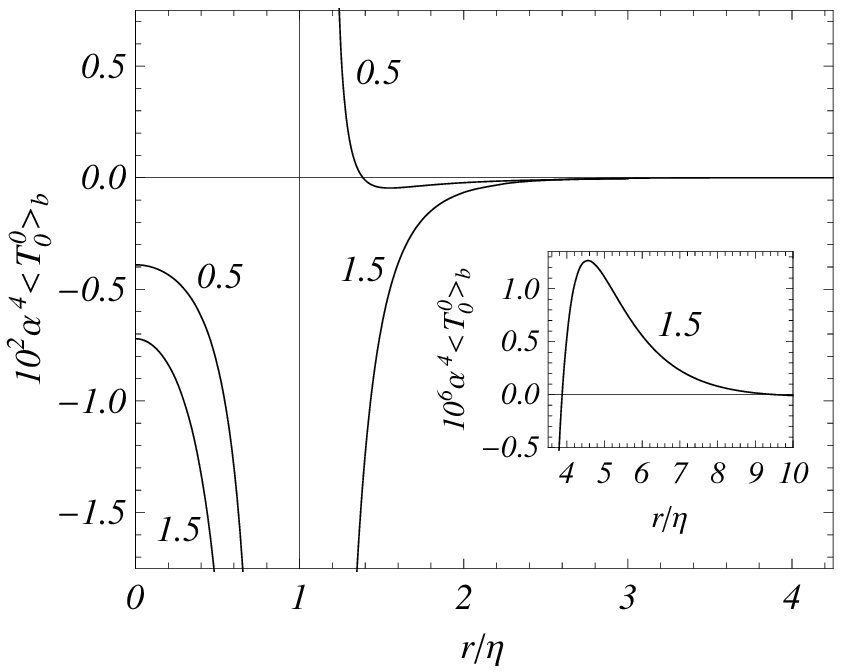,width=7.cm,height=6.cm} & \quad %
\epsfig{figure=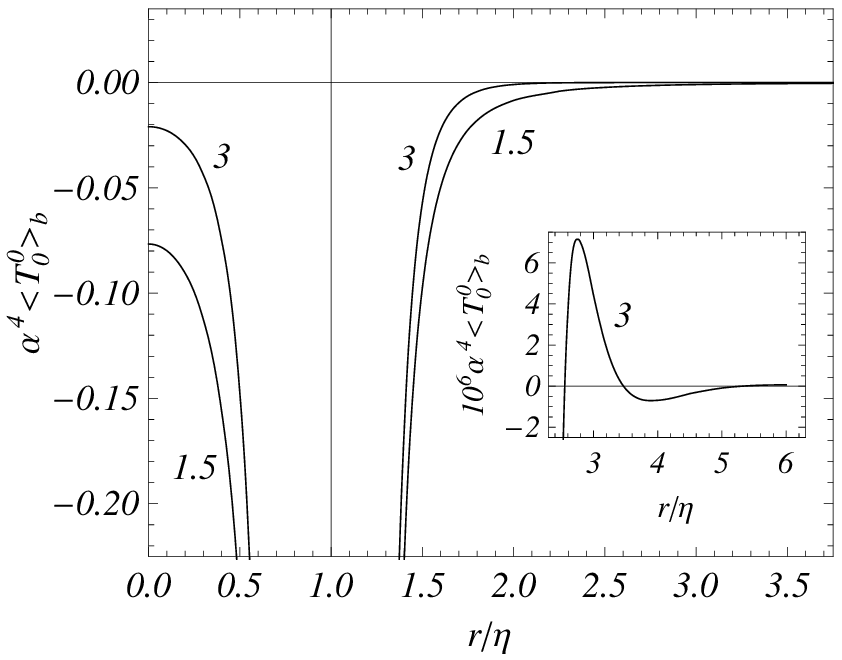,width=7.cm,height=6.cm}%
\end{tabular}%
\end{center}
\caption{The boundary-induced part in the VEV of the energy density as a
function of the proper distance from the center of the sphere for
conformally (left plot) and minimally (right plot) coupled $D=3$ scalar
fields with the Dirichlet boundary condition. The numbers near the curves
are the corresponding values of the parameter $m\protect\alpha $ and for the
sphere's proper radius we have taken $a/\protect\eta =1$.}
\label{fig1}
\end{figure}

In Fig.~\ref{fig1}, for $D=3$ dS space, we have plotted the boundary-induced
part in the VEV of the energy density for a scalar field with the Dirichlet
boundary condition as a function of the proper distance form the center of
the sphere measured in units of the dS curvature scale $\alpha $.(In figures
below we plot the VEVs for both interior and exterior regions. The
calculations for the exterior region are presented in Sect. \ref{sec:VEVex}%
.) For the sphere's proper radius (in units of $\alpha $) we have taken the
value $a/\eta =1$. The left/right panel corresponds to comformally/minimally
coupled scalar fields. The numbers near the curves are the values of the
parameter $m\alpha $. These values are chosen in a way to have cases of both
real and purely imaginary $\nu $. For a minimally coupled field the
dependence of the VEVs on the mass is weak for points near the sphere. This
follows from the asymptotic expansion (\ref{TiiNear}). For a conformally
coupled field the leading term vanishes and the subleading terms in the
asymptotic expansions depend on the mass. Note that for conformally and
minimally coupled fields the graphs have different scales. For non-Dirichlet
boundary conditions, the VEV of the energy density near the boundary has
opposite sign compared with the Dirichlet case.

\begin{figure}[tbph]
\begin{center}
\begin{tabular}{cc}
\epsfig{figure=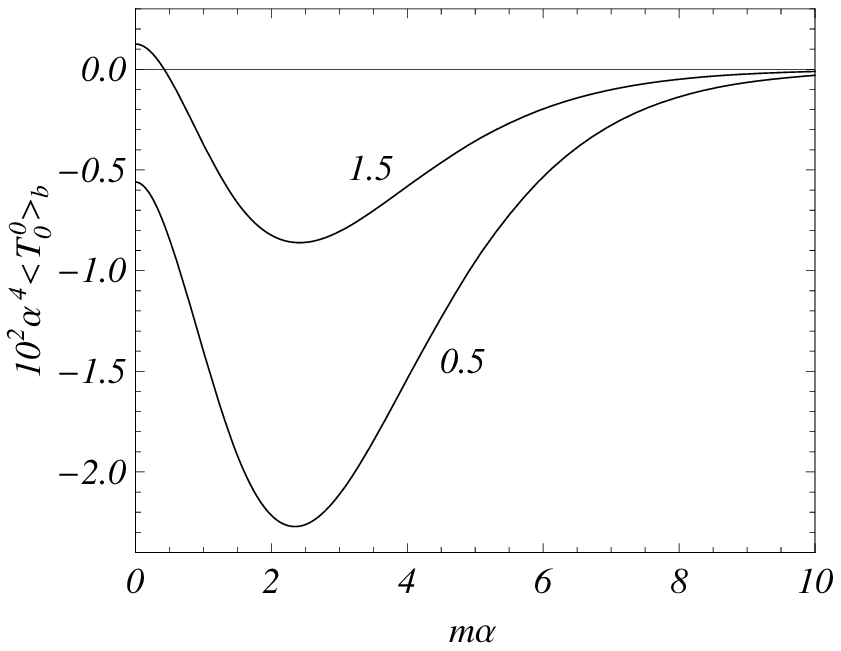,width=7.cm,height=6.cm} & \quad %
\epsfig{figure=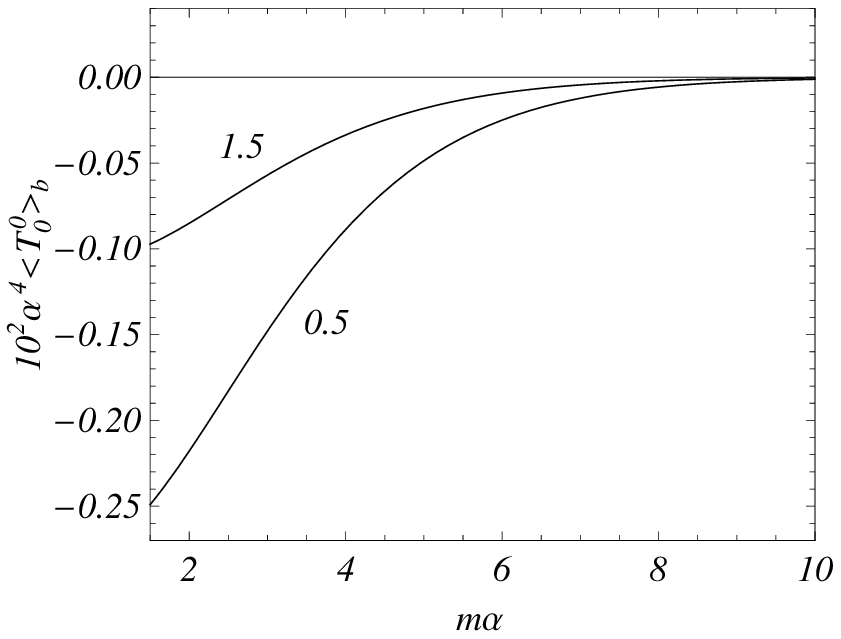,width=7.cm,height=6.cm}%
\end{tabular}%
\end{center}
\caption{The boundary-induced part in the VEV of the energy density as a
function of the field mass for a fixed proper distance from the center of
the sphere in $D=3$ dS spacetime. The left and right panels correspond to
conformally and minimally coupled fields. The numbers near the curves are
the values of the proper distance from the sphere's center in units of $%
\protect\alpha $.}
\label{fig2}
\end{figure}

Fig.~\ref{fig2} displays the dependence of the boundary-induced part in the
energy density, for a fixed proper distance from the center of the sphere
with the Dirichlet boundary condition, as a function of the mass measured in
units of the dS energy scale. As in Fig.~\ref{fig1}, the left and right
panels are for $D=3$ conformally and minimally coupled fields respectively
and the sphere radius is taken $a/\eta =1$. The numbers near the curves are
the values of the ratio $r/\eta $. For large values of $m\alpha $ the VEVs
exhibit an oscillatory behavior.

\begin{figure}[tbph]
\begin{center}
\begin{tabular}{cc}
\epsfig{figure=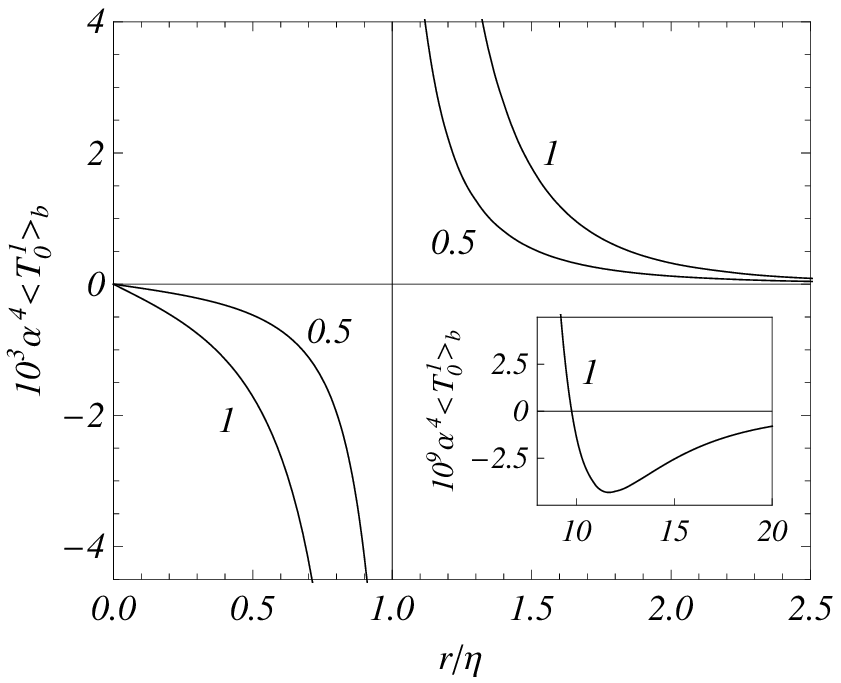,width=7.cm,height=6.cm} & \quad %
\epsfig{figure=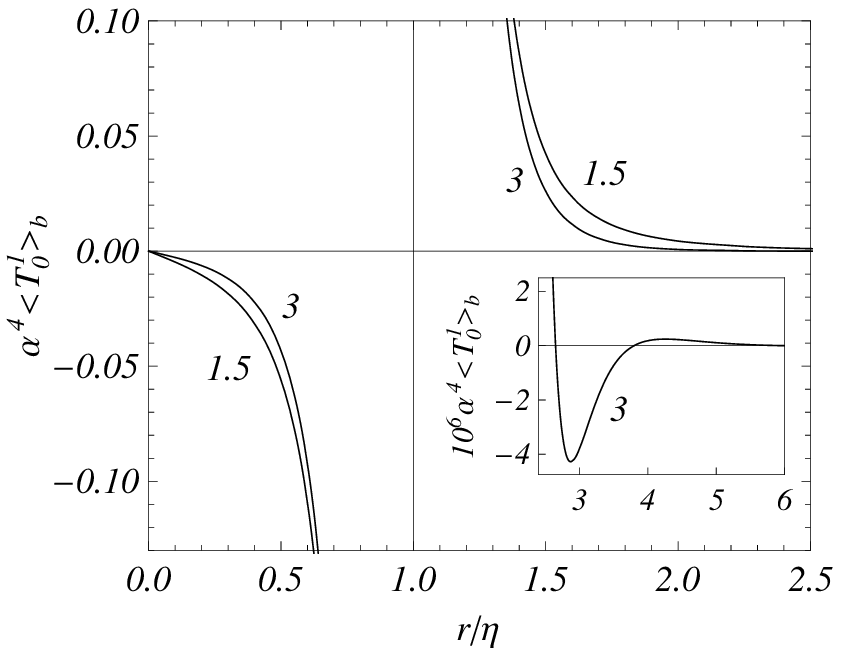,width=7.cm,height=6.cm}%
\end{tabular}%
\end{center}
\caption{The energy flux as a function of the proper distance from the
center of the sphere for $D=3$ conformally and minimally coupled scalar
fields (left and right plots respectively) with the Dirichlet boundary
condition. The numbers near the curves correspond to the values of the
parameter $m\protect\alpha $.}
\label{fig3}
\end{figure}

Fig.~\ref{fig3} shows the energy flux as a function of the proper distance
from the center of the sphere with the radius $a/\eta =1$ for $D=3$
conformally and minimally coupled fields with the Dirichlet boundary
condition. Again, for a minimally coupled scalar field the dependence on the
mass is weak, which directly follows from the asymptotic expression (\ref%
{T10bNear}). For both cases of minimally and conformally coupled fields the
energy flows out of the sphere. For non-Dirichlet boundary conditions the
energy flux has the opposite sign.

\begin{figure}[tbph]
\begin{center}
\begin{tabular}{cc}
\epsfig{figure=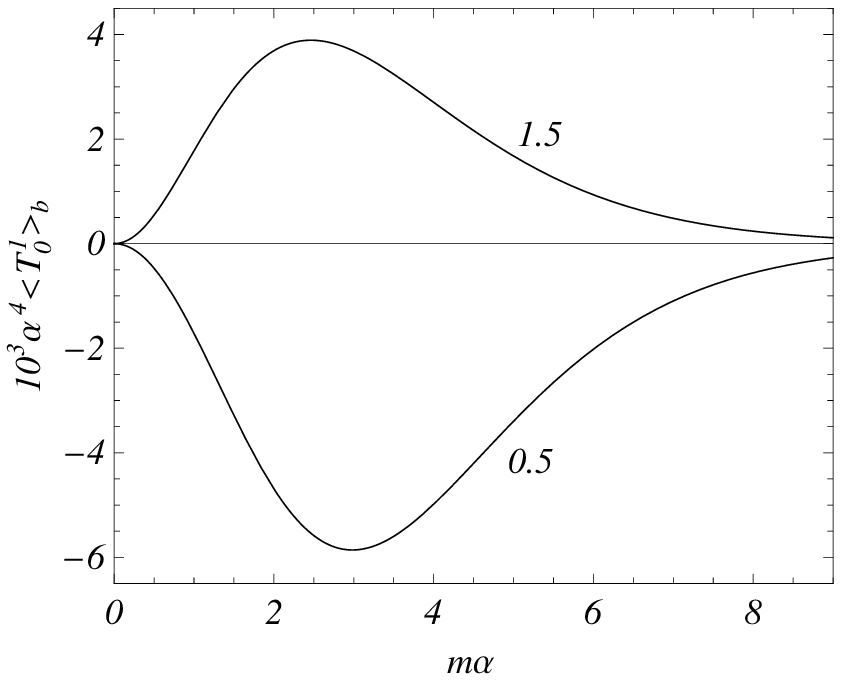,width=7.cm,height=6.cm} & \quad %
\epsfig{figure=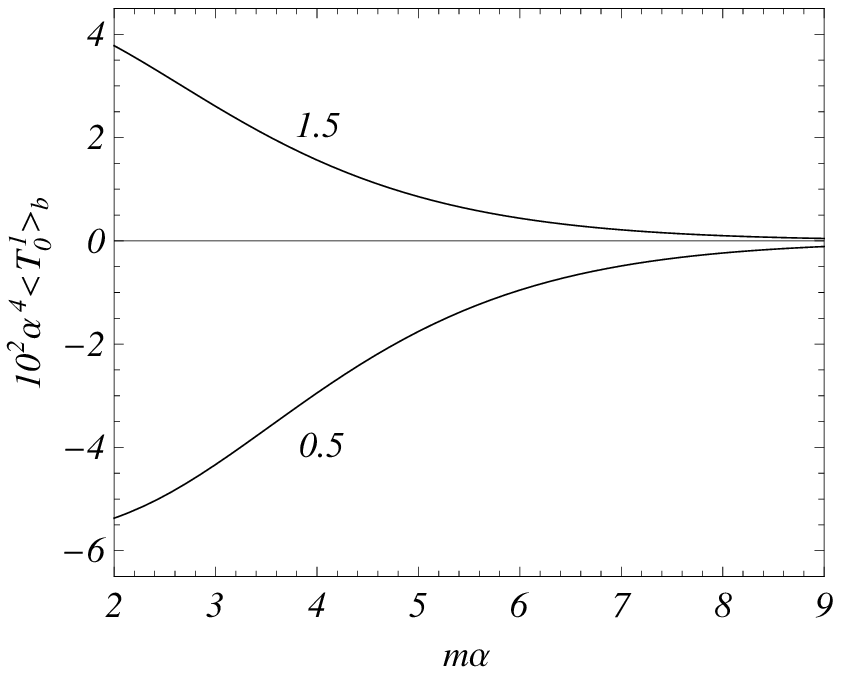,width=7.cm,height=6.cm}%
\end{tabular}%
\end{center}
\caption{The energy flux as a function of the mass for a fixed proper
distance from the center of the sphere with the proper radius $a/\protect%
\eta =1$. The left and right panels are for $D=3$ conformally and minimally
coupled fields with the Dirichlet boundary condition. The numbers near the
curves are the values of the ratio $r/\protect\eta $ (proper distance from
the sphere's center in units of $\protect\alpha $).}
\label{fig4}
\end{figure}

The dependence of the energy flux on the mass of the field is depicted in
Fig.~\ref{fig4} for a fixed value of the proper distance from the sphere's
center. As before, the case of a $D=3$ scalar field with the Dirichlet
boundary condition is considered. For a conformally coupled massless field
the energy flux vanishes

In this section, we have investigated the VEV of the bulk energy-momentum
tensor inside a spherical boundary. For a scalar fields with Robin boundary
conditions, it was shown in Ref. \cite{Rome02} that in the discussion of the
relation between the mode sum energy, evaluated as the sum of the zero-point
energies for each normal mode of frequency, and the volume integral of the
renormalized energy density it is necessary to include in the energy a
surface term concentrated on the boundary (see also Ref. \cite{Full03}). An
expression for the surface energy-momentum tensor for a scalar field with a
general curvature coupling parameter in the general case of bulk and
boundary geometries is derived in Ref.~\cite{Saha01EMT}. By making use of
this expression and the mode functions from Eq.~(\ref{eigfunc}), we can
evaluate the mode sum for the surface energy-momentum tensor. The latter is
located on the sphere and, in addition to the divergences of the
boundary-free dS spacetime, contains surface divergences. The corresponding
renormalization procedure is similar to that for the total Casimir energy
and will be discussed elsewhere.

\section{Wightman function in the exterior region}

\label{sec:Ex}

Now we turn to the evaluation of the VEVs in the region outside a spherical
shell with radius $a$. The boundary condition for a scalar field is in the
form (\ref{spherebc}). As we did for the interior region, first we consider
the Wightman function. This function is evaluated by using the mode sum
formula (\ref{WF}). In the exterior region, $r>a$, the radial parts of the
mode functions are given by Eq.~(\ref{TR}), where the ratio of the
coefficients in the linear combination of the Bessel and Neumann functions
is determined from the boundary condition (\ref{spherebc}). Note that the
boundary condition can be written in the covariant form $(1+\beta
n^{l}\nabla _{l})\varphi =0$, with $n^{l}$ being the outward normal (with
respect to the region under consideration) to the boundary normalized as $%
g_{il}n^{i}n^{l}=-1$. For the interior and exterior regions we have $%
n^{l}=(\eta /\alpha )\delta _{1}^{l}$ and $n^{l}=-(\eta /\alpha )\delta
_{1}^{l}$, respectively. Comparing with the boundary condition in the form (%
\ref{spherebc}), we see that $\tilde{B}/\tilde{A}=(\eta /\alpha )\beta $ for
the interior region and $\tilde{B}/\tilde{A}=-(\eta /\alpha )\beta $ for the
exterior one. Hence, when we impose Robin boundary condition with the same
coefficient $\beta $ then the ratio $\tilde{B}/\tilde{A}$ has opposite sign
for the exterior and interior region. In general, $\beta $ could be
different for these regions.

In the exterior region, for the mode functions realizing the Bunch-Davies
vacuum state, one finds%
\begin{equation}
\varphi _{\sigma }(x)=C_{\sigma }\frac{\eta ^{D/2}}{r^{D/2-1}}H_{\nu
}^{(1)}(\lambda \eta )g_{\mu }(\lambda a,\lambda r)Y(m_{k};\vartheta ,\phi ),
\label{EigFuncEx}
\end{equation}%
where we have defined%
\begin{equation}
g_{\mu }(\lambda a,\lambda r)=\bar{Y}_{\mu }(\lambda a)J_{\mu }(\lambda r)-%
\bar{J}_{\mu }(\lambda a)Y_{\mu }(\lambda r),  \label{gmu}
\end{equation}%
with $0\leqslant \lambda <\infty $ and with the barred notation (\ref{barnot}%
). The normalization coefficient is determined from the condition (\ref%
{Normaliz}), where now the integration goes over the exterior region and in
the right-hand side the delta symbol for $\lambda $ is understood as the
Dirac delta function $\delta (\lambda -\lambda ^{\prime })$. In evaluating
the normalization integral over the radial coordinate we note that this
integral is divergent for $\lambda =\lambda ^{\prime }$ and, hence, the
dominant contribution comes from large values of $r$. By making use of the
asymptotic formulas for the Bessel and Neumann functions for large
arguments, from Eq. (\ref{Normaliz}) one gets%
\begin{equation}
C_{\sigma }^{2}=\frac{\pi \lambda e^{i(\nu -\nu ^{\ast })\pi /2}}{4\alpha
^{D-1}N(m_{k})}\left[ \bar{J}_{\mu }^{2}(\lambda a)+\bar{Y}_{\mu
}^{2}(\lambda a)\right] ^{-1},  \label{Cext}
\end{equation}%
where we have used Eq. (\ref{HankRel}).

Substituting the functions (\ref{EigFuncEx}) into the mode sum (\ref{WF}),
for the Wightman function we find the expression:
\begin{eqnarray}
W(x,x^{\prime }) &=&\frac{\pi e^{i(\nu -\nu ^{\ast })\pi /2}}{4\alpha
^{D-1}nS_{D}}\frac{\left( \eta \eta ^{\prime }\right) ^{D/2}}{(rr^{\prime
})^{D/2-1}}\sum_{l=0}^{\infty }(2l+n)C_{l}^{n/2}(\cos \theta )  \notag \\
&&\times \int_{0}^{\infty }d\lambda \lambda \frac{g_{\mu }(\lambda a,\lambda
r)g_{\mu }(\lambda a,\lambda r^{\prime })}{\bar{J}_{\mu }^{2}(\lambda a)+%
\bar{Y}_{\mu }^{2}(\lambda a)}H_{\nu }^{(1)}(\lambda \eta )H_{\nu ^{\ast
}}^{(2)}(\lambda \eta ^{\prime }).  \label{WFex}
\end{eqnarray}%
Similar to the case for the interior region, this function can be written in
the decomposed form (\ref{WF1}). In order to extract the boundary-induced
part explicitly, we subtract from Eq.~(\ref{WFex}) the function $W_{\text{dS}%
}(x,x^{\prime })$, written in the form (\ref{WdS}). For the further
evaluation of the difference we use the relation
\begin{equation}
\frac{g_{\mu }(\lambda a,\lambda r)g_{\mu }(\lambda a,\lambda r^{\prime })}{%
\bar{J}_{\mu }^{2}(\lambda a)+\bar{Y}_{\mu }^{2}(\lambda a)}-J_{\mu
}(\lambda r)J_{\mu }(\lambda r^{\prime })=-\frac{1}{2}\sum_{j=1,2}\frac{\bar{%
J}_{\mu }(\lambda a)}{\bar{H}_{\mu }^{(j)}(\lambda a)}H_{\mu }^{(j)}(\lambda
r)H_{\mu }^{(j)}(\lambda r^{\prime }).  \label{relab}
\end{equation}%
As a result, the boundary-induced part is presented in the form%
\begin{eqnarray}
W_{\text{b}}(x,x^{\prime }) &=&-\frac{\pi e^{i(\nu -\nu ^{\ast })\pi
/2}(\eta \eta ^{\prime })^{D/2}}{8\alpha ^{D-1}nS_{D}(rr^{\prime })^{D/2-1}}%
\sum_{l=0}^{\infty }\left( 2l+n\right) C_{l}^{n/2}(\cos \theta
)\sum_{j=1,2}\int_{0}^{\infty }d\lambda  \notag \\
&&\times \,\lambda \frac{\bar{J}_{\mu }(\lambda a)}{\bar{H}_{\mu
}^{(j)}(\lambda a)}H_{\mu }^{(j)}(\lambda r)H_{\mu }^{(j)}(\lambda r^{\prime
})H_{\nu }^{(1)}(\lambda \eta )H_{\nu ^{\ast }}^{(2)}(\lambda \eta ^{\prime
}).  \label{WFbex1}
\end{eqnarray}%
Assuming that the function $\bar{H}_{\mu }^{(1)}(z)$, ($\bar{H}_{\nu
}^{(2)}(z)$) has no zeros for $0<\mathrm{arg}\,z\leqslant \pi /2$ ($-\pi
/2\leqslant \mathrm{arg}\,z<0$), we rotate the integration contour by the
angle $\pi /2$ for the term with $j=1$ and by the angle $-\pi /2$ for $j=2$.
Introducing the modified \ Bessel functions, for the boundary-induced part
of the Wightman function in the exterior region we get the expression
\begin{eqnarray}
W_{\text{b}}(x,x^{\prime }) &=&-\frac{\alpha ^{1-D}}{\pi nS_{D}}\frac{(\eta
\eta ^{\prime })^{D/2}}{(rr^{\prime })^{D/2-1}}\sum_{l=0}^{\infty }\left(
2l+n\right) C_{l}^{n/2}(\cos \theta )\int_{0}^{\infty }dx\,x  \notag \\
&&\times \frac{\bar{I}_{\mu }(ax)}{\bar{K}_{\mu }(ax)}K_{\mu }(rx)K_{\mu
}(r^{\prime }x)\left[ I_{\nu }(x\eta ^{\prime })K_{\nu }(x\eta )+I_{-\nu
}(x\eta )K_{\nu }(x\eta ^{\prime })\right] .  \label{WFbex2}
\end{eqnarray}%
Comparing with Eq. (\ref{Wb}), we see that the expressions for the boundary
induced parts in the exterior and interior regions are related by the
interchange $K_{\mu }\rightleftarrows I_{\mu }$.

\section{Vacuum expectation values in the exterior region}

\label{sec:VEVex}

\subsection{Field squared}

First we consider the VEV for the field squared. It is presented in the
decomposed form (\ref{phi2dec}), with the boundary-induced part given by%
\begin{equation}
\langle \varphi ^{2}\rangle _{\text{b}}=-\frac{\alpha ^{1-D}}{\pi
S_{D}(r/\eta )^{D-2}}\sum_{l=0}^{\infty }D_{l}\int_{0}^{\infty }dx\,x^{1-D}%
\frac{\bar{I}_{\mu }(ax/\eta )}{\bar{K}_{\mu }(ax/\eta )}K_{\mu
}^{2}(rx/\eta )F_{\nu }(x),  \label{phi2bex}
\end{equation}%
where $F_{\nu }(x)$ is defined by Eq.~(\ref{Fnu}). The time dependence in
this expression appears through the proper radius of the sphere and the
proper distance from the center. For a conformally coupled massless field
one has the relation (\ref{phi2conf}), where the expression for $\langle
\varphi ^{2}\rangle _{\text{M}}$ is obtained from Eq. (\ref{phi2M}) by the
interchange $K_{\mu }\rightleftarrows I_{\mu }$. For points near the sphere
the leading term in the asymptotic expansion over the distance from the
sphere is in the form (\ref{phi2near}) with $(a-r)$ replaced by $(r-a)$. In
particular, near the sphere the VEV has the same sign for the interior and
exterior regions.

Now we turn to the asymptotic behavior of the boundary-induced VEV at large
distances from the sphere, assuming that $r/\eta \gg 1$ for fixed $a/\eta $.
In this limit, the dominant contribution to the integral in Eq. (\ref%
{phi2bex}) comes from the region near the lower limit of the integration, $%
x\lesssim \eta /r$, and we can use the relation%
\begin{equation}
\frac{\bar{I}_{\mu }(z)}{\bar{K}_{\mu }(z)}\approx 2\frac{A+\mu B}{A-\mu B}%
\frac{(z/2)^{2\mu }}{\mu \Gamma ^{2}(\mu )},  \label{IKFas}
\end{equation}%
for $z\ll 1$. By using also Eq.~(\ref{KIas}), the integral in Eq.~(\ref%
{phi2bex}) involving the square of the Macdonald function is evaluated with
the help of a formula from Ref.~\cite{Prud86}. The dominant contribution
comes from the mode with $l=0$. The behavior of the boundary-induced VEV at
large distances is qualitatively different for positive and imaginary values
of the parameter $\nu $. In the first case, the leading term is given by the
expression%
\begin{eqnarray}
\langle \varphi ^{2}\rangle _{\text{b}} &\approx &-\frac{2\alpha ^{1-D}}{\pi
S_{D}}\frac{\Gamma (\nu )\Gamma \left( n/4+1-\nu \right) }{n\Gamma
^{2}(n/2)\Gamma (1-\nu )}  \notag \\
&&\times \Gamma \left( 3n/4+1-\nu \right) \frac{A+nB/2}{A-nB/2}\frac{(a/\eta
)^{D-2}}{(r/\eta )^{2D-2\nu -2}},  \label{phi2bexLarg1}
\end{eqnarray}%
and the VEV is a monotonically decreasing function of the radial coordinate.

At large distances from the sphere and for imaginary $\nu $, $\nu =i|\nu |$,
the VEV behaves as%
\begin{eqnarray}
\langle \varphi ^{2}\rangle _{\text{b}} &\approx &-\frac{C_{0}(\nu )\alpha
^{1-D}}{\pi ^{D/2+1}\Gamma (D/2-1)}\frac{A+nB/2}{A-nB/2}  \notag \\
&&\times \frac{(a/\eta )^{D-2}}{(r/\eta )^{2D-2}}\cos [2|\nu |\ln (r/\eta
)+\phi _{0}],  \label{phi2bexLarg2}
\end{eqnarray}%
where the coefficient $C_{0}$ and the phase $\phi _{0}$ are defined by the
relation%
\begin{equation}
C_{0}(\nu )e^{i\phi _{0}}=\frac{\Gamma (\nu )}{\Gamma (1-\nu )}\Gamma \left(
3n/4-\nu +1\right) \Gamma \left( n/4-\nu +1\right) .  \label{C0}
\end{equation}%
In particular, for $D=3$ one has%
\begin{equation}
C_{0}(\nu )=\pi \sqrt{2}\frac{|3/16-\nu +\nu ^{2}|}{|\nu |\cosh ^{1/2}(2\pi
|\nu |)}.  \label{C0D3}
\end{equation}%
As is seen, for imaginary $\nu $ the behavior of the boundary-induced VEV\
is damped oscillatory. For a fixed value of $r$, the VEV behaves as $\langle
\varphi ^{2}\rangle _{\text{b}}\sim e^{-Dt/\alpha }\cos (2|\nu |t/\alpha
+\phi _{0}^{\prime })$, where $\phi _{0}^{\prime }=2|\nu |\ln (r/\alpha
)+\phi _{0}$. The oscillation frequency is increasing with increasing mass
of the field.

It is of interest to compare the behavior of the VEV at large distances to
the corresponding behavior for a spherical boundary in Minkowski spacetime.
In the latter case, for a massless field the decay is of power-law: $\langle
\varphi ^{2}\rangle _{\text{M}}\sim a^{D-2}/r^{2D-3}$. For a massive field,
assuming that $mr\gg 1$, the VEV behaves as%
\begin{equation}
\langle \varphi ^{2}\rangle _{\text{M}}\approx -\frac{\sqrt{\pi }}{%
4S_{D}r^{D-1}}\frac{e^{-2mr}}{\sqrt{mr}}\sum_{l=0}^{\infty }D_{l}\frac{\bar{I%
}_{\mu }(am)}{\bar{K}_{\mu }(am)},  \label{phi2Mlarg}
\end{equation}%
and it is exponentially suppressed. In the case of dS spacetime, for a fixed
value of $m\alpha \lesssim 1$, the decay at large distances is of power law
(monotonic or oscillatory) for both massless and massive fields. So, we
conclude that the curvature of the background spacetime essentially changes
the behavior of the VEVs at distances larger than the curvature radius of
the background spacetime.

\subsection{VEV of the energy-momentum tensor}

Now we turn to the VEV of the energy-momentum tensor outside the sphere.
This VEV is presented in the decomposed form (\ref{TikDec}). By using the
formula (\ref{emtvev1}) and Eq. (\ref{phi2bex}) for the VEV of the field
squared, the diagonal components of the boundary-induced part are presented
in the form (no summation over $i$)%
\begin{eqnarray}
\langle T_{i}^{i}\rangle _{\text{b}} &=&-\frac{r^{2-D}\eta ^{2}}{2\pi
S_{D}\alpha ^{D+1}}\sum_{l=0}^{\infty }D_{l}\int_{0}^{\infty }dx\,x^{3-D}%
\frac{\bar{I}_{\mu }(ax)}{\bar{K}_{\mu }(ax)}  \notag \\
&&\times \left\{ G_{i}^{i}[K_{\mu }(rx)]F_{\nu }(x\eta )+2K_{\mu
}^{2}(rx)F_{i}^{i}(x\eta )\right\} ,  \label{Tiibex}
\end{eqnarray}%
where the functions $F_{i}^{i}(y)$ and $G_{i}^{i}\left[ f(y)\right] $ are
defined by Eqs.~(\ref{Fii}) and (\ref{Gii}). For the off-diagonal component
we have the expression%
\begin{eqnarray}
\langle T_{0}^{1}\rangle _{\text{b}} &=&\frac{\eta \alpha ^{-D-1}}{\pi
S_{D}r^{D-1}}\sum_{l=0}^{\infty }D_{l}\int_{0}^{\infty }dx\,x^{1-D}\frac{%
\bar{I}_{\mu }(ax)}{\bar{K}_{\mu }(ax)}K_{\mu }(rx)  \notag \\
&&\times \left[ 2rxK_{\mu }^{\prime }(rx)-(D-2)K_{\mu }(rx)\right] \left[
\left( \xi -1/4\right) \eta \partial _{\eta }+\xi \right] F_{\nu }(x\eta ).
\label{T01bex}
\end{eqnarray}%
These components are written in the coordinate system $(\tau ,r,\vartheta
,\phi )$. They are related to the components in the system $(t,r,\vartheta
,\phi )$ by Eq. (\ref{ConfSyncrel}).

The vacuum energy in the region $r\geqslant r_{0}>a$ is given by the
expression (\ref{EbV}) with the integration over this region. Making use of
the expression for the energy density from Eq. (\ref{Tiibex}), one finds the
formula%
\begin{eqnarray}
E_{r\geqslant r_{0}}^{\text{(b)}} &=&\frac{(r_{0}/\eta )^{2-D}}{2\pi \alpha }%
\sum_{l=0}^{\infty }D_{l}\int_{0}^{\infty }dx\,x^{1-D}\frac{\bar{I}_{\mu
}(ax/r_{0})}{\bar{K}_{\mu }(ax/r_{0})}  \notag \\
&&\times \big\{\left( 1-4\xi \right) F_{\nu }(x\eta /r_{0})K_{\mu }(x)\left[
x\,K_{\mu }^{\prime }(x)-\frac{D-2}{2}K_{\mu }(x)\right]  \notag \\
&&-F_{0}^{0}(x\eta /r_{0})\left[ x^{2}K_{\mu }^{\prime 2}(x)-(x^{2}+\mu
^{2})K_{\mu }^{2}(x)\right] \big\}.  \label{EtotEx}
\end{eqnarray}%
This energy is related to the energy flux and to the vacuum stresses in the
exterior region by Eq. (\ref{EnCons}). Similar to the interior region, the
integrated energy diverges in the limit $r_{0}\rightarrow a$.

Let us consider the asymptotics for the VEV of the energy-momentum tensor.
For points near the sphere, in a way similar to that for the interior region
we find for $r\rightarrow a$ (no summation over $k$)%
\begin{eqnarray}
\langle T_{k}^{k}\rangle _{\text{b}} &\approx &\frac{D\Gamma ((D+1)/2)(\xi
-\xi _{D})\kappa _{B}}{2^{D}\pi ^{(D+1)/2}[\alpha (r-a)/\eta ]^{D+1}},
\notag \\
\langle T_{1}^{1}\rangle _{\text{b}} &\approx &-\frac{(D-1)\Gamma
((D+1)/2)(\xi -\xi _{D})\kappa _{B}}{2^{D}\pi ^{(D+1)/2}a(r-a)^{D}(\alpha
/\eta )^{D+1}},  \label{TiiNearex}
\end{eqnarray}%
for $k=0,2,\ldots ,D$, and
\begin{equation}
\langle T_{0}^{1}\rangle _{\text{b}}\approx -\frac{D\Gamma (D/2)\left( \xi
-\xi _{D}\right) \kappa _{B}}{2^{D}\pi ^{D/2+1}\alpha \lbrack \alpha
(r-a)/\eta ]^{D}}.  \label{T10bNearex}
\end{equation}%
As we see, for a scalar field with non-conformal coupling the components $%
\langle T_{k}^{k}\rangle _{\text{b}}$, $k=0,2,\ldots ,D$, have the same sign
near the sphere for exterior and interior regions, whereas the radial stress
and the off-diagonal component change signs.

Now we consider the limit of large distances from the sphere, assuming that $%
r/\eta \gg 1$ for fixed $a/\eta $. Similar to the case of the field squared,
discussed in the previous subsection, for positive values of $\nu $ to the
leading order we get:%
\begin{equation}
\langle T_{i}^{k}\rangle _{\text{b}}\approx \frac{(\eta /r)^{D-2\nu
+1-\delta _{i}^{k}}(a/r)^{D-2}}{2\pi ^{D/2+1}\Gamma (D/2-1)\alpha ^{D+1}}%
\frac{\tilde{A\,}B_{i}^{k}(\nu )}{\tilde{A}+\left( D-2\right) \tilde{B}/a}.
\label{TikexLarg}
\end{equation}%
In this expression,%
\begin{eqnarray}
B_{0}^{0}(\nu ) &=&\frac{D\Gamma (\nu )}{\Gamma (1-\nu )}\left[ \left(
D+1-2\nu \right) \xi -\frac{D-2\nu }{4}\right]  \notag \\
&&\times \Gamma \left( 3n/4+1-\nu \right) \Gamma \left( n/4+1-\nu \right) ,
\label{B00}
\end{eqnarray}%
and (no summation over $k$)%
\begin{eqnarray}
B_{k}^{k}(\nu ) &=&\frac{2\nu }{D}B_{0}^{0}(\nu ),\;k=1,2,\ldots ,D,  \notag
\\
B_{0}^{1}(\nu ) &=&-2\frac{D-1-\nu }{D}B_{0}^{0}(\nu ).  \label{B01}
\end{eqnarray}%
In this case the boundary-induced VEV decays monotonically with increasing
proper distance from the sphere. At large distances the vacuum stresses are
isotropic. If we denote the corresponding effective pressure by $P_{\text{b}%
} $, $P_{\text{b}}=-\langle T_{1}^{1}\rangle _{\text{b}}$, then the equation
of state is of the barotropic type: $P_{\text{b}}=-(2\nu /D)\langle
T_{0}^{0}\rangle _{\text{b}}$.

At large distances from the sphere and for imaginary $\nu $, the leading
term in the asymptotic expansion is in the form%
\begin{equation}
\langle T_{i}^{k}\rangle _{\text{b}}\approx \frac{(\eta /r)^{D+1-\delta
_{i}^{k}}(a/r)^{D-2}C_{i}^{k}(\nu )}{\pi ^{D/2+1}\Gamma (D/2-1)\alpha ^{D+1}}%
\frac{\widetilde{A\,}\cos [2|\nu |\ln (r/\eta )+\phi _{i}^{k}]}{\widetilde{A}%
+\left( D-2\right) \widetilde{B}/a},  \label{TikexLargIm}
\end{equation}%
where $C_{i}^{k}(\nu )$ and the phases $\phi _{i}^{k}$ are defined by the
relation%
\begin{equation}
B_{i}^{k}(\nu )=C_{i}^{k}(\nu )e^{i\phi _{i}^{k}}.  \label{BikIm}
\end{equation}%
As in the previous case, to the leading order the vacuum stresses are
isotropic. The oscillations in the energy density and the stresses are
shifted by the phase $\pi /2$. For a fixed comoving radial distance $r$, the
VEVs oscillate like $\langle T_{i}^{k}\rangle _{\text{b}}\sim
e^{-(D+1-\delta _{i}^{k})t/\alpha }\cos (2|\nu |t/\alpha +\phi _{i}^{k\prime
})$. For the Neumann boundary condition, the leading terms, given by Eqs.~(%
\ref{TikexLarg}) and (\ref{TikexLargIm}), vanish and it is necessary to take
the next terms in the asymptotic expansions. As a result the asymptotic
expressions for the Neumann boundary condition contain an additional factor $%
(a/r)^{2}$ compared with non-Neumann boundary conditions. In Figs.~\ref{fig1}%
--\ref{fig4}, the boundary-induced part of the energy density and the energy
flux in the exterior region are depicted as functions of the proper distance
from the sphere center and of the mass, for $D=3$ minimally and conformally
coupled fields with the Dirichlet boundary condition and for the sphere's
proper radius corresponding to $a/\eta =1$. Both cases of real and imaginary
$\nu $ are considered. The oscillatory behavior at large distances in the
latter case is plotted separately in the inset. The first zero decreases
with increasing mass, whereas the oscillation frequency increases.

\section{Further generalizations}

\label{sec:Gen}

The approach described above may be used for the evaluation of the Casimir
densities for more general background geometries, in particular, in models
where the spherical boundary separates spacetime regions with different
geometries. Here, as an example, we consider a model in which the spacetime
is described by two distinct metric tensors in the regions outside and
inside the spherical boundary with radius $a$. The line element in the
exterior region will be taken in the form (\ref{ds2Dd}). We will assume that
inside the sphere the spacetime geometry is regular and is described by the
general spherically symmetric line element%
\begin{equation}
ds^{2}=\alpha ^{2}\tau
^{-2}[e^{2u(r)}dt^{2}-e^{2v(r)}dr^{2}-e^{2w(r)}d\Omega _{D}^{2}],
\label{ds2inside}
\end{equation}%
where the functions $u(r)$, $v(r)$, $w(r)$ are continuous at the core
boundary:
\begin{equation}
u(a)=v(a)=0,\;w(a)=\ln (a).  \label{uvbound}
\end{equation}%
Here we assume that there is no surface energy-momentum tensor located at $%
r=a$ and, hence, the derivatives of these functions are continuous as well.
Note that by introducing the new radial coordinate $\tilde{r}=e^{w(r)}$ with
the sphere's center at $\tilde{r}=0$, the angular part of the line element (%
\ref{ds2inside}) is written in the standard Minkowskian form. With this
coordinate, in general, we will obtain a non-standard angular part in the
exterior line element (\ref{ds2Dd}). From the regularity of the interior
geometry at the sphere center one has the conditions $u(r),v(r)\rightarrow 0$%
, and $w(r)\sim \ln \tilde{r}$ for $\tilde{r}\rightarrow 0$.

The curvature scalar for the metric corresponding to (\ref{ds2inside}) is
presented in the form%
\begin{equation}
R=\left( \alpha /\eta \right) ^{-2}\check{R}+D(D+1)\alpha ^{-2}e^{-2u(r)},
\label{Rgen}
\end{equation}%
where%
\begin{eqnarray}
\check{R} &=&-2e^{-2v}\left[ u^{\prime \prime }+u^{\prime 2}-u^{\prime
}v^{\prime }+(D-2)(D-1)w^{\prime 2}/2\right.   \notag \\
&&\left. +(D-1)\left( w^{\prime \prime }+w^{\prime 2}+w^{\prime }u^{\prime
}-w^{\prime }v^{\prime }\right) \right] +(D-2)(D-1)e^{-2w}  \label{Rstat}
\end{eqnarray}%
is the curvature scalar for the static metric corresponding to the
line element inside the square brackets in (\ref{ds2inside}).
Explicitly writing the field equation (\ref{fieldeq}) for the
metric (\ref{ds2inside}), it can be seen that the time and radial
variables are separated in two special cases: for $u(r)=0$ or for
a massless field. We will consider the first case. The discussion
for the second case is similar.

Taking $u(r)=0$ and assuming the Bunch-Davies vacuum state, the mode
functions in the region inside the sphere, $r<a$, are written in the form%
\begin{equation}
\varphi _{\sigma }(x)=\eta ^{D/2}H_{\nu }^{(1)}(\lambda \eta
)f_{l}(r)Y(m_{p};\vartheta ,\phi ),  \label{ModeFuncin}
\end{equation}%
where the radial function is a solution of the equation%
\begin{equation}
f_{l}^{\prime \prime }(r)+[(D-1)w^{\prime }-v^{\prime }]f_{l}^{\prime
}(r)+e^{2v}\left[ \lambda ^{2}-\xi \check{R}-l(l+D-2)e^{-2w}\right]
f_{l}(r)=0.  \label{RadEq}
\end{equation}%
and%
\begin{equation}
\check{R}=-2(D-1)e^{-2v}\left( w^{\prime \prime }+Dw^{\prime 2}/2-w^{\prime
}v^{\prime }\right) +(D-1)(D-2)e^{-2w}.  \label{RstatSp}
\end{equation}

The solution of the radial equation (\ref{RadEq}) regular at the origin we
will denote by $R_{l}(r,\lambda )$. Near the center this solution behaves
like $\tilde{r}^{l}$. Note that the parameter $\lambda $ enters in the
radial equation in the form $\lambda ^{2}$. As a result the regular solution
can be chosen in such a way that $R_{l}(r,-\lambda )=\mathrm{const}\cdot
R_{l}(r,\lambda )$. Now for the radial part of the eigenfunctions one has%
\begin{equation}
f_{l}(r)=\left\{
\begin{array}{ll}
R_{l}(r,\lambda ) & \mathrm{for}\;r<a \\
r^{1-D/2}\left[ A_{l}J_{\mu }(\lambda r)+B_{l}Y_{\mu }(\lambda r)\right] &
\mathrm{for}\;r>a%
\end{array}%
.\right.  \label{fltot}
\end{equation}%
The coefficients $A_{l}$ and $B_{l}$ are determined by the conditions of
continuity of the radial function and its derivative at $r=a$:
\begin{eqnarray}
A_{l} &=&\frac{\pi }{2}a^{D/2-1}R_{l}(a,\lambda )\check{Y}_{\mu }(\lambda
a),\;  \notag \\
B_{l} &=&-\frac{\pi }{2}a^{D/2-1}R_{l}(a,\lambda )\check{J}_{\mu }(\lambda
a),  \label{AlBl}
\end{eqnarray}%
with the notation, for a cylinder function $F(z)$,
\begin{equation}
\check{F}(z)\equiv zF^{\prime }(z)-\left[ \frac{D}{2}-1+a\frac{R_{l}^{\prime
}(a,z/a)}{R_{l}(a,z/a)}\right] F(z),  \label{Hatnot}
\end{equation}%
where $R_{l}^{\prime }(a,\lambda )=\partial _{r}R_{l}(r,\lambda )|_{r=a}$.
Due to our choice of the function $R_{l}(r,\lambda )$, the logarithmic
derivative in formula (\ref{Hatnot}) is an even function of $z$. Hence, in
the region $r>a$ the radial part of the mode functions becomes%
\begin{equation}
f_{l}(r)=\frac{\pi }{2}(a/r)^{D/2-1}R_{l}(a,\lambda )h_{\mu }(\lambda
a,\lambda r),  \label{fl2}
\end{equation}%
with the notation
\begin{equation}
h_{\mu }(\lambda a,\lambda r)=\check{Y}_{\mu }(\lambda a)J_{\mu }(\lambda r)-%
\check{J}_{\mu }(\lambda a)Y_{\mu }(\lambda r).  \label{hnot}
\end{equation}

The normalization condition (\ref{Normaliz}) is written in terms of the
radial eigenfunctions as
\begin{equation}
\int_{r_{0}}^{\infty }dr\sqrt{|g_{r}|}f_{l}(r,\lambda )f_{l}(r,\lambda
^{\prime })=\frac{\pi \delta (\lambda -\lambda ^{\prime })}{4N(m_{k})\alpha
^{D-1}}e^{i(\nu -\nu ^{\ast })\pi /2},  \label{normfl}
\end{equation}%
where $r_{0}$ is the value of the radial coordinate $r$ corresponding to the
sphere center and $g_{r}$ is the radial part of the determinant $g$. As the
integral on the left is divergent for $\lambda ^{\prime }=\lambda $, the
main contribution in the coincidence limit comes from large values $r$. \ By
using the expression (\ref{fl2}) for the radial part in the region $r>a$ and
replacing the Bessel and Neumann functions by the leading terms of their
asymptotic expansions for large values of the argument, it can be seen that
from (\ref{normfl}) the following result is obtained:%
\begin{equation}
R_{l}^{2}(a,\lambda )=\frac{a^{2-D}\alpha ^{1-D}\lambda e^{i(\nu -\nu ^{\ast
})\pi /2}}{\pi N(m_{k})\left[ \check{J}_{\mu }^{2}(\lambda a)+\check{Y}_{\mu
}^{2}(\lambda a)\right] }.  \label{normcoefRl}
\end{equation}

By taking into account relation (\ref{normcoefRl}), the mode functions in
the exterior region are presented in the form%
\begin{equation}
\varphi _{\sigma }(x)=c_{\sigma }\frac{\eta ^{D/2}}{r^{D/2-1}}H_{\nu
}^{(1)}(\lambda \eta )h_{\mu }(\lambda a,\lambda r)Y(m_{k};\vartheta ,\phi ),
\label{ModeExtGen}
\end{equation}%
where%
\begin{equation}
c_{\sigma }^{2}=\frac{\pi \lambda e^{i(\nu -\nu ^{\ast })\pi /2}}{4\alpha
^{D-1}N(m_{k})}\left[ \check{J}_{\mu }^{2}(\lambda a)+\check{Y}_{\mu
}^{2}(\lambda a)\right] ^{-1}.  \label{NormCoefGen}
\end{equation}%
Now comparing with (\ref{EigFuncEx}), we see that the mode functions (\ref%
{ModeExtGen}) are obtained from the corresponding functions for the sphere
with Robin boundary condition by the replacement%
\begin{equation}
\frac{A}{B}\rightarrow -\frac{D}{2}+1-a\frac{R_{l}^{\prime }(a,\lambda )}{%
R_{l}(a,\lambda )}.  \label{ReplGen}
\end{equation}%
As a result, the Wightman function in the exterior region is given by
expression (\ref{WFex}) with this replacement.

Further evaluation of the Wightman function and the Casimir
densities is the same as in the case of the Robin sphere,
described in Sections \ref{sec:Ex} and \ref{sec:VEVex}. The
Wightman function in the region $r>a$ is presented in the
decomposed form (\ref{WF1}), where now the part induced by the
geometry (\ref{ds2inside}) with $u(r)=0$ in the region $r<a$ is given by (%
\ref{WFbex2}) with
\begin{equation}
\frac{A}{B}=-\frac{D}{2}+1-a\frac{R_{l}^{\prime }(a,xe^{\pi i/2})}{%
R_{l}(a,xe^{\pi i/2})}.  \label{ABGen}
\end{equation}%
The factor $e^{\pi i/2}$ in the argument of the function $R_{l}$
is related to the complex rotation we have used after formula
(\ref{WFbex1}). Similarly, the expressions for corresponding parts
in the VEVs of the field
squared and the energy-momentum tensor are presented as (\ref{phi2bex}) and (%
\ref{Tiibex}) with the ratio of the coefficients from (\ref{ABGen}).

\section{Conclusion}

\label{sec:Conc}

In this paper, we have investigated the Casimir densities induced by a
spherical boundary in dS spacetime for a massive scalar field with general
curvature coupling parameter. On the sphere the field obeys Robin boundary
condition with coefficients, in general, different for the interior and
exterior regions. We have assumed that the field is prepared in the
Bunch-Davies vacuum state which is dS-invariant in the boundary-free
spacetime. In free field theories all properties of a quantum field are
encoded in two-point functions. We have computed the Wightman function and
the Green's function both in the region interior to the sphere and in its
exterior. The VEVs for the field squared and the energy-momentum tensor are
obtained from these two-point functions in the coincidence limit. In
addition, the Wightman function determines the response of particle
detectors of the Unruh-DeWitt type. In a similar way other two-point
functions can be investigated.

For the evaluation of the Wightman function we have employed the mode
summation method. In the region inside the spherical boundary, the mode
functions for a scalar field, realizing the Bunch-Davies vacuum state, are
given by Eq.~(\ref{eigfunc}). The eigenvalues for $\lambda $ are quantized
by the boundary condition on the sphere and they are the solutions of Eq.~(%
\ref{eigenmodes}). The corresponding mode sum for the Wightman function is
given by Eq.~(\ref{WF0}) and contains the summation over these eigenvalues.
The latter are given implicitly and Eq. (\ref{WF0}) is not convenient for
the further evaluation of the VEVs in the coincidence limit. In order to
obtain a more workable form, we have used the generalized Abel-Plana formula
for the summation over the eigenvalues of $\lambda $. This allowed us (i) to
extract explicitly the boundary-free Wightman function and (ii) to present
the sphere-induced part in terms of an integral rapidly convergent in the
coincidence limit (for points away from the boundary). The local geometry
away from the sphere is the same as in the boundary-free dS spacetime and
the renormalization for the VEVs of the field squared and the
energy-momentum tensor is reduced to that for the boundary-free dS
spacetime. The latter is well investigated in the literature. In the same
way we can evaluate the Wightman function in a general vacuum state
corresponding to the mode functions (\ref{TR}) with a linear combination of
Hankel functions. The corresponding expression for the boundary-induced part
has the form similar to Eq.~(\ref{Wb}) with a more general bilinear
combination of the modified Bessel functions of order $\nu $.

By using the representation of the Wightman function, the VEVs of the field
squared and the energy-momentum tensor are decomposed into the boundary-free
and boundary-induced parts. For the region inside the sphere the latter are
given by Eqs.~(\ref{phi2b}), (\ref{Tiib}), and (\ref{T01b}), for the field
squared and the energy-momentum tensor, respectively. An important feature
is that the vacuum energy-momentum tensor has an off-diagonal component (\ref%
{T01b}) which describes energy flux along the radial direction. With
dependence on the boundary condition and the mass of the field, this flux
can be either positive or negative. The boundary induced VEVs for both field
squared and the energy-momentum tensor depend on time through the proper
radius of the sphere and the proper distance from the sphere's center. This
property is a consequence of the maximal symmetry of the Bunch-Davies vacuum
state.

We have explicitly checked that the boundary-induced part in the VEV of the
energy-momentum tensor obeys the trace relation (\ref{TrRel}) and the
covariant conservation equation. In particular, the energy-momentum tensor
is traceless for a conformally coupled massless field. The trace anomaly is
present in the boundary-free part only. For a conformally coupled massless
scalar field the flux vanishes. In this case we have simple relations, Eqs. (%
\ref{phi2conf}) and (\ref{Tiiconf}), between the boundary-induced VEVs for
spherical boundaries in dS and Minkowski spacetimes. The latter are
consequences of the conformal relation between the problems in Minkowski and
dS spacetimes. Divergences are found in the VEVs as the surface of the
sphere is approached. The leading terms in the corresponding asymptotic
expansions in terms of the distance from the boundary are given by Eqs. (\ref%
{phi2near}), (\ref{TiiNear}) and (\ref{T10bNear}). Written in terms of the
proper distance from the boundary, for a non-conformally coupled field these
leading terms coincide with the corresponding terms for a spherical boundary
in Minkowski spacetime. This is a consequence of the fact that for points
near the boundary the dominant contribution to the VEVs comes from the modes
with small wavelengths which are not influenced by the gravitational field.

In the region outside a spherical shell, the eigenvalues for the quantum
number $\lambda $ are continuous and the modes of the field realizing the
Bunch-Davies vacuum state are given by Eq.~(\ref{EigFuncEx}). We have
explicitly extracted from the Wightman function the part corresponding to dS
spacetime without boundaries. The boundary-induced part is given by Eq.~(\ref%
{WFbex2}) for the Wightman function and by Eqs.~(\ref{phi2bex}) and (\ref%
{Tiibex}), (\ref{T01bex}) for the field squared and the energy-momentum
tensor, respectively. General formulas are simplified in the asymptotic
regions. For points near the boundary the leading terms in the expansions
over the distance from the sphere are given by expressions (\ref{TiiNearex})
and (\ref{T10bNearex}). In this region the energy density and the azimuthal
stresses have the same sign for the exterior and interior regions, whereas
the radial stress and the energy flux have opposite signs. In particular,
for a minimally coupled scalar field the energy flows away from the boundary
for Dirichlet boundary condition and toward the boundary for non-Dirichlet
boundary conditions.

Most interesting is the behavior far from the sphere; qualitatively
different behavior occurs depending on the sign of $\nu ^{2}=D^{2}/2-\xi
D(D+1)-m^{2}\alpha ^{2}$, where $D$ is the number of spatial dimensions, $%
\xi $ is the conformal parameter, $m$ is the mass of the scalar field, and $%
\alpha $ is the curvature scale. When $\nu $ is positive, the mean field
squared and the energy-momentum tensor fall off as a power, while when $\nu $
is imaginary, the large distance behavior is damped oscillatory with the
amplitude decaying as $(\eta /r)^{2(D-1)}$ for the field squared and the
diagonal components of the energy-momentum tensor. For a scalar field with
Neumann boundary condition the VEVs at large distances are suppressed by an
additional factor $(\eta /r)^{2}$ compared with the case of non-Neumann
boundary conditions. Note that the behavior of the VEVs at distances larger
than the curvatures scale of the background spacetime is completely
different from the case of a spherical boundary in Minkowski spacetime. In
the latter case and for a massive field the boundary-induced VEVs at
distances larger than the Compton wavelength of the scalar particle are
exponentially suppressed by the factor $e^{-m(r-a)}$. For the problem in dS
spacetime, under the condition $m\lesssim \alpha ^{-1}$, the decay of the
VEVs is power-law. Exponential damping can occur only in an intermediate
region $\alpha \gg r\gg 1/m$.

In Section \ref{sec:Gen} we have generalized the expressions of the Casimir
densities in the exterior region for a class of spherically-symmetric
metrics in the region $r<a$ described by the line element (\ref{ds2inside}).
The geometry in the exterior region is given by the dS line element (\ref%
{ds2Dd}). A special case with $u(r)=0$ is considered when the time and
radial variables in the field equation are separated. The VEVs in the
exterior region are decomposed as the sum of pure dS part and the part
induced by the geometry in the interior region. We have shown that the
expressions for the latter are obtained from the corresponding expressions
outside the Robin sphere, investigated in Sections \ref{sec:Ex} and \ref%
{sec:VEVex}, taking the ratio of the Robin coefficients in the form (\ref%
{ABGen}), where $R_{l}(r,\lambda )$ is the solution of the interior radial
equation regular at the origin.

At the end, we would like to emphasize that the main subject of
the present paper is the investigation of the local
characteristics of the vacuum, the VEVs for the field squared and
the energy-momentum tensor, at the points away from the
boundaries. They do not contain surface divergences and are
completely determined within the framework of standard
renormalization procedure in quantum field theory without
boundaries. We expect that similar results would be obtained in
the model in which instead of externally imposed boundary
condition the fluctuating field is coupled to a smooth background
potential that implements the boundary condition in a certain limit \cite%
{Grah02}.

\section*{Acknowledgments}

This work was supported in part by grants from the US National
Science Foundation and the US Department of Energy. The visit of
A.A.S. to the University of Oklahoma was supported in part by an
International Travel Grant from the American Physical Society.
A.A.S. is grateful to the Homer L. Dodge Department of Physics and
Astronomy and the University of Oklahoma for their kind
hospitality.

\appendix

\section{Boundary-free part of the Wightman function}

\label{sec:AppBF}

For the further evaluation of the boundary-free part, given by Eq. (\ref{WdS}%
), firstly we apply Gegenbauer's addition theorem for the Bessel functions
(see, for instance, Ref.~\cite{Abra72}) to the series over $l$. This gives
the following expression for the Wightman function:%
\begin{eqnarray}
W_{\text{dS}}(x,x^{\prime }) &=&\frac{\pi e^{i(\nu -\nu ^{\ast })\pi /2}}{%
\Gamma (n/2)nS_{D}}\frac{2^{-n/2-1}\alpha ^{1-D}\left( \eta \eta ^{\prime
}\right) ^{D/2}}{\left( r^{2}+r^{\prime 2}-2rr^{\prime }\cos \theta \right)
^{n/4}}  \notag \\
&&\times \int_{0}^{\infty }dz\,z^{n/2+1}H_{\nu }^{(1)}(z\eta )H_{\nu ^{\ast
}}^{(2)}(z\eta ^{\prime })J_{n/2}(z\sqrt{r^{2}+r^{\prime 2}-2rr^{\prime
}\cos \theta }).  \label{WFdS1}
\end{eqnarray}%
As the next step, we write the product of the Hankel functions in terms of
the Macdonald function:%
\begin{equation}
e^{i(\nu -\nu ^{\ast })\pi /2}H_{\nu }^{(1)}(z\eta )H_{\nu ^{\ast
}}^{(2)}(z\eta ^{\prime })=\frac{4}{\pi ^{2}}K_{\nu }(-iz\eta )K_{\nu
}(iz\eta ^{\prime }),  \label{HH}
\end{equation}%
and use the integral representation \cite{Watson}%
\begin{equation}
K_{\nu }(Z)K_{\nu }(z)=\frac{1}{4}\int_{-\infty }^{+\infty
}dy\int_{0}^{\infty }\frac{dw}{w}e^{-\nu y-Zzw^{-1}\cosh y}\exp \left( -%
\frac{w}{2}-\frac{Z^{2}+z^{2}}{2w}\right)  \label{KK}
\end{equation}%
for the product of the Macdonald functions. Substituting Eqs.~(\ref{HH}) and
(\ref{KK}) into Eq.~(\ref{WFdS1}), we first integrate over $x$ and then with
respect to $w$, with the result%
\begin{equation}
W_{\text{dS}}(x,x^{\prime })=\frac{\alpha ^{1-D}}{2\pi S_{D}}%
\int_{0}^{\infty }dz\frac{z^{\nu +D/2-1}}{[z^{2}-2u(x,x^{\prime })z+1]^{D/2}}%
,  \label{WFdS2}
\end{equation}%
where
\begin{equation}
u(x,x^{\prime })=1+\frac{(\eta -\eta ^{\prime })^{2}-r^{2}-r^{\prime
2}+2rr^{\prime }\cos \theta }{2\eta \eta ^{\prime }}.  \label{u}
\end{equation}%
In deriving Eq. (\ref{WFdS2}) we have assumed that $|u|<1$. The integral in
Eq.~(\ref{WFdS2}) is expressed in terms of the associated Legendre function $%
P_{\nu -1/2}^{(1-D)/2}(u(x,x^{\prime }))$ (see Ref.~\cite{GradShtein}).
Expressing this function through the hypergeometric function, after some
transformations, we get the final expression for the Wightman function in dS
spacetime (for two-point functions in boundary-free dS spacetime see Ref.~%
\cite{Cand75}):%
\begin{eqnarray}
W_{\text{dS}}(x,x^{\prime }) &=&\frac{\alpha ^{1-D}}{(4\pi )^{(D+1)/2}}\frac{%
\Gamma (D/2+\nu )\Gamma (D/2-\nu )\,}{\Gamma ((D+1)/2)}  \notag \\
&&\times \,_{2}F_{1}\left( \frac{D}{2}+\nu ,\frac{D}{2}-v;\frac{D+1}{2};%
\frac{1+u(x,x^{\prime })}{2}\right) .  \label{WFdS3}
\end{eqnarray}%
Note that, if we denote by $X(x)$ the coordinates in the higher-dimensional
embedding space for dS spacetime, then one can write $u(x,x^{\prime
})=1+[X(x)-X(x^{\prime })]^{2}/(2\alpha ^{2})$. The property that the
Wightman function depends on spacetime points through $[X(x)-X(x^{\prime
})]^{2}$ is related to the maximal symmetry of the Bunch-Davies vacuum state.

\section{Minkowski spacetime limit}

\label{sec:AppMink}

In this appendix, for the boundary-induced part in the Wightman function, we
explicitly demonstrate the limiting transition to the geometry of a
spherical boundary in Minkowski spacetime. In the Minkowski spacetime limit
one has $\alpha \rightarrow \infty $ and the modulus of the order of the
modified Bessel functions in Eq.~(\ref{WF1}) is large, $\nu \approx i\sigma $%
, $\sigma =m\alpha \gg 1$. In addition, we have $\eta \approx \alpha -t$. We
make use of the uniform asymptotic expansions for the modified Bessel
functions for imaginary values of the order with large modulus. For $z<1$,
the leading terms in these expansions have the form (see, for example, Ref.~%
\cite{Milt09})%
\begin{eqnarray}
&&K_{i\sigma }(\sigma z)\sim \sqrt{\frac{2\pi }{\sigma }}e^{-\sigma \pi
/2}\cos [\sigma f(z)-\pi /4],  \notag \\
&&I_{i\sigma }(\sigma z)+I_{-i\sigma }(\sigma z)\sim -\frac{2e^{\sigma \pi
/2}}{\sqrt{2\pi \sigma }}\sin [\sigma f(z)-\pi /4],  \label{IKas1}
\end{eqnarray}%
where
\begin{equation}
f(z)=\ln \left( \frac{1+\sqrt{1-z^{2}}}{z}\right) -\sqrt{1-z^{2}}.
\label{fzas}
\end{equation}%
In the case $z>1$ one has the asymptotics
\begin{eqnarray}
&&K_{i\sigma }(\sigma z)\sim \sqrt{\frac{\pi }{2\sigma }}\frac{e^{-\sigma
\pi /2-\sigma g(z)}}{(z^{2}-1)^{1/4}},  \notag \\
&&I_{i\sigma }(\sigma z)+I_{-i\sigma }(\sigma z)\sim \frac{2}{\sqrt{2\pi
\sigma }}\frac{e^{\sigma \pi /2+\sigma g(z)}}{(z^{2}-1)^{1/4}},
\label{IKas2}
\end{eqnarray}%
with
\begin{equation}
g(z)=-{\mathrm{arcsec\,}}z+\sqrt{z^{2}-1},\quad g^{\prime }(z)=\frac{1}{z}%
\sqrt{z^{2}-1}.  \label{gz}
\end{equation}

From Eqs.~(\ref{IKas1}) and (\ref{IKas2}) it follows that the dominant
contribution to the boundary-induced part of the Wightman function in Eq.~(%
\ref{Wb}) comes from the integration range $x>m$. In this range we have
\begin{equation}
I_{\nu }(x\eta ^{\prime })K_{\nu }(x\eta )+I_{-\nu }(x\eta )K_{\nu }(x\eta
^{\prime })\approx \frac{\cosh \left\{ m\alpha \left[ g(z)-g(z^{\prime })%
\right] \right\} }{m\alpha (z^{\prime 2}-1)^{1/4}(z^{2}-1)^{1/4}},
\label{IKas}
\end{equation}%
where $z=x\eta /m\alpha $ and $z^{\prime }=x\eta ^{\prime }/m\alpha $. By
taking into account that $\eta /\alpha \approx 1-t/\alpha $ and $\eta
^{\prime }/\alpha \approx 1-t^{\prime }/\alpha $, it can be seen that%
\begin{equation}
g(z)-g(z^{\prime })\approx \sqrt{\left( x/m\right) ^{2}-1}\left( t^{\prime
}-t\right) /\alpha .  \label{gzz}
\end{equation}%
Now, combining this with Eq. (\ref{IKas}), we get%
\begin{equation}
I_{\nu }(x\eta ^{\prime })K_{\nu }(x\eta )+I_{-\nu }(x\eta )K_{\nu }(x\eta
^{\prime })\approx \frac{\cosh (\Delta t\sqrt{x^{2}-m^{2}})}{\alpha \sqrt{%
x^{2}-m^{2}}},  \label{IKas3}
\end{equation}%
with $\Delta t=t^{\prime }-t$. Substituting this into the expression for the
boundary-induced part of the Wightman function, Eq. (\ref{Wb}), to the
leading order one finds
\begin{eqnarray}
W_{\text{b}}(x,x^{\prime }) &\approx &-\frac{(rr^{\prime })^{-n/2}}{\pi
nS_{D}}\sum_{l=0}^{\infty }(2l+n)C_{l}^{n/2}(\cos \theta )\int_{m}^{\infty
}\ dx\,x  \notag \\
&&\times \frac{\bar{K}_{\mu }(xa)}{\bar{I}_{\mu }(xa)}\frac{\cosh (\Delta t%
\sqrt{x^{2}-m^{2}})}{\sqrt{x^{2}-m^{2}}}I_{\mu }(xr)I_{\mu }(xr^{\prime }).
\label{WFM}
\end{eqnarray}%
The expression in the right-hand side coincides with the Wightman function
inside a spherical boundary in the Minkowski bulk \cite{Saha01}. The
Minkowski space limit for the Wightman function in the exterior region is
considered in a similar way.

\section{Green's function}

\label{app:C}

In the text, we computed the Wightman function in order to calculate the
VEVs of the field-squared and the energy-momentum tensor. Of course, these
quantities can equally well be computed from the Green's function for this
problem. The calculation of the latter is similar to that given for the
Wightman function, but there are some points of interest, so we sketch the
derivation here.

Because the nontrivial structure in the dS geometry lies in the time
dependence, it is natural to isolate that by constructing a reduced Green's
function in the variable $\eta $, $\eta ^{\prime }$ ($n=D-2$):
\begin{equation}
G(x,x^{\prime })=\frac{\alpha ^{1-D}}{nS_{D}(rr^{\prime })^{n/2}}%
\sum_{l=0}^{\infty }(2l+n)C_{l}^{n/2}(\cos \theta )\sum_{\lambda }\frac{%
2\lambda }{a}T_{\mu }(\lambda a)J_{\mu }(\lambda r)J_{\mu }(\lambda
r^{\prime })f(\eta ,\eta ^{\prime }).  \label{GF}
\end{equation}%
Here the notation is as in Sec.~\ref{sec:WF}, and in particular, the
eigenvalues $\lambda $ are the roots of Eq.~(\ref{eigenmodes}). The reduced
Green's functions satisfies
\begin{equation}
\left[ \frac{\partial ^{2}}{\partial \eta ^{2}}+\frac{1-D}{\eta }\frac{%
\partial }{\partial \eta }+\frac{\alpha ^{2}m^{2}+\xi D(D-1)}{\eta ^{2}}%
+\lambda ^{2}\right] f(\eta ,\eta ^{\prime })=-\left( \frac{\eta }{\alpha }%
\right) ^{D-1}\delta (\eta -\eta ^{\prime }).  \label{GFeq}
\end{equation}%
The solution of this equation corresponding to the Bunch-Davies vacuum (the
analog of outgoing wave solutions) is
\begin{equation}
f(\eta ,\eta ^{\prime })=\frac{\pi }{4i}e^{i\pi (\nu -\nu ^{\ast })/2}(\eta
\eta ^{\prime })^{D/2}H_{\nu }^{(1)}(\lambda \eta _{>})H_{\nu ^{\ast
}}^{(2)}(\kappa \eta _{<}),  \label{GFred}
\end{equation}%
which uses the Wronskian (\ref{HankRel}) and $\eta _{>}$ ($\eta _{<}$) is
the greater (lesser) of $\eta $, $\eta ^{\prime }$.

To resolve the difficulties with oscillatory integrals and an implicit
equation for the eigenvalues, we can again use the generalized Abel-Plana
formula (\ref{sumJ1anal}), which leads to the breakup of the Green's
function into a free part referring only to the dS background, and a term
which exhibits the effect of the sphere:
\begin{equation}
G(x,x^{\prime })=G_{\text{dS}}(x,x^{\prime })+G_{\text{b}}(x,x^{\prime }),
\label{GFdec}
\end{equation}%
where
\begin{eqnarray}
G_{\text{dS}}(x,x^{\prime }) &=&-\frac{\pi }{4i}\frac{e^{i\pi (\nu -\nu
^{\ast })/2}}{\alpha ^{D-1}nS_{D}}\frac{(\eta \eta ^{\prime })^{D/2}}{%
(rr^{\prime })^{n/2}}\sum_{l=0}^{\infty }(2l+n)C_{l}^{n/2}(\cos \theta )
\notag \\
&&\quad \times \int_{0}^{\infty }d\lambda \,\lambda \,J_{\mu }(\lambda
r)J_{\mu }(\lambda r^{\prime })H_{\nu }^{(1)}(\kappa r_{>})H_{\nu ^{\ast
}}^{(2)}(\lambda r_{<}),  \label{GF1}
\end{eqnarray}%
and
\begin{eqnarray}
G_{\text{b}}(x,x^{\prime }) &=&\frac{1}{\pi i}\frac{\alpha ^{1-D}}{nS_{D}}%
\frac{(\eta \eta ^{\prime })^{D/2}}{(rr^{\prime })^{n/2}}\sum_{l=0}^{\infty
}(2l+n)C_{l}^{n/2}(\cos \theta )  \notag \\
&&\times \int_{0}^{\infty }d\lambda \,\lambda \,\frac{\bar{K}_{\mu }(\lambda
a)}{\bar{I}_{\mu }(\lambda a)}I_{\mu }(\lambda r)I_{\mu }(\lambda r^{\prime
})\left[ K_{\nu }(\lambda \eta _{>})I_{-\nu }(\lambda \eta _{<})+K_{\nu
}(\lambda \eta _{<})I_{\nu }(\lambda \eta _{>})\right] .  \label{GFb}
\end{eqnarray}%
There is, in fact, no discontinuity associated with the last factor, since
the quantity in square brackets here is
\begin{equation}
\frac{\pi /2}{\sin \pi \nu }\left[ I_{-\nu }(\lambda \eta )I_{-\nu }(\lambda
\eta ^{\prime })-I_{\nu }(\lambda \eta )I_{\nu }(\lambda \eta ^{\prime })%
\right] .  \label{rel3}
\end{equation}%
Of course, $G_{\text{b}}(x,x^{\prime })$ is simply $iW_{\text{b}%
}(x,x^{\prime })$ given in Eq.~(\ref{Wb}).

The calculation of the Green's function for the exterior region proceeds
similarly. Again, if we impose the boundary condition at the smallest value
of $r$ in the region, here at $r=a$, we have the radial function $g_{\mu }$
given in Eq.~(\ref{gmu}), which have continuum normalization
\begin{equation}
\int_{a}^{\infty }dr\,r\,g_{\mu }(\lambda r)g_{\mu }(\lambda r^{\prime })=%
\frac{1}{\lambda }\delta (\lambda -\lambda ^{\prime })[\bar{J}_{\mu
}^{2}(\lambda a)+\bar{Y}_{\mu }^{2}(\lambda a)].  \label{NormInt}
\end{equation}%
Thus the exterior Green's function is
\begin{eqnarray}
G(x,x^{\prime }) &=&\frac{i\pi }{4}\frac{e^{i\pi (\nu -\nu ^{\ast })/2}}{%
\alpha ^{D-1}nS_{D}}\sum_{l=0}^{\infty }(2l+D-2)C_{l}^{n/2}(\cos \theta )
\notag \\
&& \times \int_{0}^{\infty }d\lambda \frac{\lambda g_{\mu }(\lambda r)g_{\mu
}(\lambda r^{\prime })}{\bar{J}_{\mu }^{2}(\lambda a)+\bar{Y}_{\mu
}^{2}(\lambda a)}H_{\nu }^{(1)}(\lambda \eta _{>})H_{\nu ^{\ast
}}^{(2)}(\lambda \eta _{<}).  \label{GFext}
\end{eqnarray}%
This evidently leads to the boundary-dependent part given by Eq.~(\ref%
{WFbex2}) multiplied by $i$.

Incidentally, we might record here the imaginary rotations of the Hankel
functions, since these are not given in the standard tables:
\begin{eqnarray}
H_{\nu ^{\ast }}^{(2)}(e^{\pi i/2}x) &=&2\left[ e^{i\nu ^{\ast }\pi
/2}I_{\nu ^{\ast }}(x)+\frac{i}{\pi }e^{-i\nu ^{\ast }\pi /2}K_{\nu }(x)%
\right] ,  \notag \\
H_{\nu }^{(1)}(e^{-\pi i/2}x) &=&2\left[ e^{-i\nu \pi /2}I_{\nu ^{\ast }}(x)-%
\frac{i}{\pi }e^{i\nu \pi /2}K_{\nu }(x)\right] .  \label{Hk}
\end{eqnarray}%
for real or imaginary $\nu $, which terms are related by evident complex
conjugation.


\begin{thebibliography}{99}
\bibitem{Lind90} A.D. Linde, \textit{Particle Physics and Inflationary
Cosmology} (Harwood Academic Publishers, Chur, Switzerland 1990).

\bibitem{Ries07} A.G. Riess et al., Astrophys. J. \textbf{659}, 98 (2007);
D.N. Spergel et al., Astrophys. J. Suppl. Ser. \textbf{170}, 377 (2007); E.
Komatsu et al., Astrophys. J. Suppl. Ser. \textbf{180}, 330 (2009).

\bibitem{Stro01} A. Strominger, J. High Energy Phys. 10(2001)034; A.
Strominger, J. High Energy Phys. 11(2001)049.

\bibitem{Most97} E. Elizalde, S.D. Odintsov, A. Romeo, A.A. Bytsenko, and S.
Zerbini, \textit{Zeta Regularization Techniques with Applications} (World
Scientific, Singapore, 1994); V.M. Mostepanenko and N.N. Trunov, \textit{The
Casimir Effect and Its Applications} (Clarendon, Oxford, 1997); K.A. Milton,
\textit{The Casimir Effect: Physical Manifestation of Zero-Point Energy}
(World Scientific, Singapore, 2002); M. Bordag, G.L. Klimchitskaya, U.
Mohideen, \ and V.M. Mostepanenko, \textit{Advances in the Casimir Effect}
(Oxford University Press, Oxford, 2009); G.L. Klimchitskaya, U. Mohideen,
and V.M. Mostepanenko, Rev. Mod. Phys. \textbf{81}, 1827 (2009).

\bibitem{Casimir53} H.B.G. Casimir, Physica \textbf{19}, 846 (1953).

\bibitem{Boyer} T.H. Boyer, Phys. Rev. \textbf{174}, 1764 (1968).

\bibitem{DaviesSph} B. Davies, J. Math. Phys. \textbf{13}, 1324 (1972); R.
Balian and B. Duplantier, Ann. Phys. (N.Y.) \textbf{112}, 165 (1978); K.A.
Milton, L.L. DeRaad, Jr., and J. Schwinger, Ann. Phys. (N. Y.) \textbf{115},
388 (1978).

\bibitem{Rome95} A. Romeo, Phys. Rev. D \textbf{52}, 7308 (1995); S.
Leseduarte and A. Romeo, Ann. Phys. \textbf{250}, 448 (1996); M. Bordag, E.
Elizalde, and K. Kirsten, J. Math. Phys. \textbf{37}, 895 (1996); J.S.
Dowker, Class. Quantum Grav. \textbf{13}, 1 (1996); M. Bordag, E. Elizalde,
K. Kirsten, and S. Leseduarte, Phys. Rev. D \textbf{56}, 4896 (1997 ); V.V.
Nesterenko and I.G. Pirozhenko, Phys. Rev. D \textbf{57}, 1284 (1998); E.
Elizalde, M. Bordag, and K. Kirsten, J. Phys. A: Math. Gen. \textbf{31},
1743 (1998); M.E. Bowers and C.R. Hagen, Phys. Rev. D \textbf{59}, 025007
(1999); G. Lambiase, V.V. Nesterenko, and M. Bordag, J. Math. Phys. \textbf{%
40}, 6254 (1999).

\bibitem{Scha98} M. Schaden and L. Spruch, Phys. Rev. A \textbf{58}, 935
(1998); M. Schaden and L. Spruch, Phys. Rev. Lett. \textbf{84}, 459 (2000);
M. Schaden, arXiv:1006.3262.

\bibitem{Jaff04} R.L. Jaffe and A. Scardicchio, Phys. Rev. Lett. \textbf{92}%
, 070402 (2004); A. Scardicchio and R.L. Jaffe, Nucl. Phys. B \textbf{704},
552 (2005); A. Scardicchio and R.L. Jaffe, Nucl.Phys. B \textbf{743}, 249
(2006).

\bibitem{Gies03} H. Gies, K. Langfeld, and L. Moyaerts, J. High Energy Phys.
\textbf{06} (2003) 018; H. Gies and K. Klingmuller, Phys. Rev. Lett. \textbf{%
96}, 220401 (2006); H. Gies and K. Klingmuller, Phys. Rev. D \textbf{74},
045002 (2006).

\bibitem{Bord85} M. Bordag, D. Robaschik, and E. Wieczorek, Ann. Phys.
\textbf{165}, 192 (1985); D. Robaschik, K. Scharnhorst, and E. Wieczorek,
Ann. Phys. \textbf{174}, 401 (1987); R. Golestanian and M. Kardar, Phys.
Rev. Lett. \textbf{78}, 3421 (1997); R. Golestanian and M. Kardar, Phys.
Rev. A \textbf{58}, 1713 (1998); T. Emig, A. Hanke, R. Golestanian, and M.
Kardar, Phys. Rev. Lett. \textbf{87}, 260402 (2001); T. Emig, A. Hanke, R.
Golestanian, and M. Kardar, Phys. Rev. A \textbf{67}, 022114 (2003); R. B%
\"{u}scher and T. Emig, Phys. Rev. Lett. \textbf{94}, 133901 (2005).

\bibitem{Gene03} C. Genet, A. Lambrecht, and S. Reynaud, Phys. Rev. A
\textbf{67}, 043811 (2003); A. Lambrecht, P.A. Maia Neto, and S. Reynaud,
New J. Phys. \textbf{8}, 243 (2006); O. Kenneth and I. Klich, Phys. Rev.
Lett. \textbf{97}, 160401 (2006); T. Emig, N. Graham, R.L. Jaffe, and M.
Kardar, Phys. Rev. Lett. \textbf{99}, 170403 (2007); T. Emig, N. Graham, R.
L. Jaffe, and M. Kardar, Phys. Rev. D \textbf{77}, 025005 (2008); K.A.
Milton and J. Wagner, J. Phys. A \textbf{41}, 155402 (2008); O. Kenneth and
I. Klich, Phys. Rev. B \textbf{78}, 014103 (2008); P. A. Maia Neto, A.
Lambrecht, and S. Reynaud, Phys. Rev. A \textbf{78}, 012115 (2008); A.
Lambrecht and V.N. Marachevsky, Phys. Rev. Lett. \textbf{101}, 160403
(2008); S. J. Rahi, T. Emig, N. Graham, R.L. Jaffe, and M. Kardar, Phys.
Rev. D \textbf{80}, 085021 (2009); S.J. Rahi, T. Emig, N. Graham, R.L.
Jaffe, and M. Kardar, Phys. Rev. D \textbf{80}, 085021 (2009).

\bibitem{Rodr07} A.W. Rodriguez, M. Ibanescu, D. Iannuzzi, J.D.
Joannopoulos, and S.G. Johnson, Phys. Rev. A \textbf{76}, 032106 (2007); M.
T. Homer Reid, A.W. Rodriguez, J. White, and S.G. Johnson, \ Phys. Rev.
Lett. \textbf{103}, 040401 (2009).

\bibitem{Lect11} \textit{Lecture Notes in Physics: Casimir Physics, }Vol.
834,\textit{\ }Eds.~Diego Dalvit, Peter Milonni, David Roberts, and Felipe
da Rosa (Springer, Berlin, 2011).

\bibitem{MiltonSc} C.M. Bender and K.A. Milton, Phys. Rev. D \textbf{50},
6547 (1994).

\bibitem{MiltonVec} K.A. Milton, Phys. Rev. D \textbf{55}, 4940 (1997).

\bibitem{Cognola} E. Cognola, E. Elizalde, and K. Kirsten, J. Phys. A
\textbf{34}, 7311 (2001).

\bibitem{Teo10} L.P. Teo, Phys. Rev. D \textbf{82}, 085009 (2010).

\bibitem{Teo11} L.P. Teo, Phys. Lett. B \textbf{696}, 529 (2011).

\bibitem{Olaussen1} K. Olaussen and F. Ravndal, Nucl. Phys. B\textbf{\ 192},
237 (1981); K. Olaussen and F. Ravndal, Phys. Lett. B \textbf{100}, 497
(1981).

\bibitem{Brevik1} I. Brevik and H. Kolbenstvedt, Ann. Phys. (N.Y.) \textbf{%
149}, 237 (1983); I. Brevik and H. Kolbenstvedt, Can. J. Phys. \textbf{62},
805 (1984).

\bibitem{Grig1} L.Sh. Grigoryan and A.A. Saharian, Dokl. Akad. Nauk Arm. SSR
\textbf{83}, 28 (1986) (in Russian); L.Sh. Grigoryan and A.A. Saharian, Izv.
Akad. Nauk. Arm. SSR Fiz. \textbf{22}, 3 (1987) [J. Contemp. Phys. \textbf{22%
}, 1 (1987)].

\bibitem{Sahrev} A.A. Saharian, \textquotedblright The Generalized
Abel-Plana Formula. Applications to Bessel Functions and Casimir
Effect\textquotedblright , Report No. IC/2000/14; hep-th/0002239.

\bibitem{dewitt} B.S. DeWitt, Phys.\ Rep.\ \textbf{19}, 295 (1975).

\bibitem{Saha01} A.A. Saharian, Phys. Rev. D \textbf{63}, 125007 (2001).

\bibitem{GlobMon} A.A. Saharian and M.R. Setare, Class. Quantum Grav.
\textbf{20}, 3765 (2003); A.A. Saharian and M.R. Setare, Int. J. Mod. Phys.
A \textbf{19}, 4301 (2004); A.A. Saharian, Astrophys. \textbf{47}, 260
(2004); A.A. Saharian and E. R. Bezerra de Mello, J. Phys. A \textbf{37},
3543 (2004); E.R. Bezerra de Mello and A. A. Saharian, Class. Quantum Grav.
\textbf{23}, 4673 (2006); E.R. Bezerra de Mello and A.A. Saharian, J. High
Energy Phys. 10(2006)049, E.R. Bezerra de Mello and A.A. Saharian, Phys.
Rev. D \textbf{75}, 065019 (2007).

\bibitem{RindSph} A.A. Saharian and M.R. Setare, Nucl. Phys. B \textbf{724},
406 (2005); A.A. Saharian and M.R. Setare, Phys. Lett. B \textbf{637}, 5
(2006); A.A. Saharian and M.R. Setare, J. High Energy Phys. 02(2007)089.

\bibitem{Seta01} M.R. Setare and R. Mansouri, Class. Quantum Grav. \textbf{18%
}, 2331 (2001); M.R. Setare, Class. Quant. Grav. \textbf{18}, 4823 (2001).

\bibitem{Saha09} A.A. Saharian and T.A. Vardanyan, Class. Quantum Grav.
\textbf{26}, 195004 (2009); E. Elizalde, A.A. Saharian, and T.A. Vardanyan,
Phys. Rev. D \textbf{81}, 124003 (2010); A.A. Saharian, arXiv:1106.1873, to
appear in Int. J. Mod. Phys. A.

\bibitem{Burd11} P. Burda, arXiv:1101.2624.

\bibitem{miltonrev10}
K. A. Milton, arXiv:1005.0031, \textit{Lecture Notes in Physics: Casimir
Physics, }Vol. 834, Eds.~Diego Dalvit, Peter Milonni, David Roberts, and
Felipe da Rosa (Springer, Berlin, 2011), pp.~39--91.

\bibitem{Saha08} A.A. Saharian and M.R. Setare, Phys. Lett. B \textbf{659},
367 (2008); S. Bellucci and A.A. Saharian, Phys. Rev. D \textbf{77}, 124010
(2008); A.A. Saharian, Class. Quantum Grav. \textbf{25}, 165012 (2008); E.R.
Bezerra de Mello and A.A. Saharian, J. High Energy Phys. \textbf{12}%
(2008)081.

\bibitem{Birr82} N.D. Birrell and P.C.W. Davies, \textit{Quantum Fields in
Curved Space} (Cambridge University Press, Cambridge, 1982).

\bibitem{Most85} V.M. Mostepanenko and N.N. Trunov, Sov. J. Nucl. Phys.
\textbf{42}, 812 (1985); S.L. Lebedev, JETP \textbf{83}, 423 (1996); S.L.
Lebedev, Phys. At. Nucl. \textbf{64}, 1337 (2001).

\bibitem{Saha08Neg} A.A. Saharian, J. Phys. A \textbf{41}, 415203 (2008);
A.A. Saharian, J. Phys. A \textbf{42}, 465210 (2009).

\bibitem{Erdelyi} A. Erd\'{e}yi \textit{et al}. \textit{Higher
Transcendental Functions}. Vol. 2 (McGraw Hill, New York, 1953).

\bibitem{Abra72} \textit{Handbook of Mathematical Functions}, edited by M.
Abramowitz and I.A. Stegun (Dover, New York, 1972).

\bibitem{Alle85} B. Allen, Phys. Rev. D \textbf{32}, 3136 (1985); B. Allen
and A. Folacci, Phys. Rev. D \textbf{35}, 3771 (1987).

\bibitem{Bunc78} T.S. Bunch and P.C.W. Davies, Proc. R. Soc. London A
\textbf{360}, 117 (1978).

\bibitem{Watson} G.N. Watson, \textit{A Treatise on the Theory of Bessel
Function} (Cambridge University Press, Cambridge, 1995).

\bibitem{Ford77} L.H. Ford and L. Parker, Phys. Rev. D \textbf{16}, 245
(1977).

\bibitem{Saha08b} A.A. Saharian, \textit{The Generalized Abel-Plana Formula
with Applications to Bessel Functions and Casimir Effect} (Yerevan State
University Publishing House, Yerevan, 2008); Preprint ICTP/2007/082;
arXiv:0708.1187.

\bibitem{estrada}
R. Estrada, S.A. Fulling, L. Kaplan, K. Kirsten, Z. Liu, and K.A. Milton, J.
Phys. A: Math. Theor. \textbf{41}, 164055 (2008).

\bibitem{fall1} 
K.A. Milton, S.A. Fulling, P. Parashar, A. Romeo, K.V. Shajesh, and J.
Wagner, 
Phys. Rev. D \textbf{76}, 025004 (2007).

\bibitem{fall2}
K.A. Milton, P. Parashar, K. V. Shajesh, and J. Wagner,
J. Phys. A: Math. Theor. \textbf{40}, 10935 (2007).

\bibitem{Cand75} P. Candelas and D.J. Raine, Phys. Rev. D \textbf{12}, 965
(1975); J.S. Dowker and R. Critchley, Phys. Rev. D \textbf{13}, 224 (1976);
J.S. Dowker and R. Critchley, Phys. Rev. D \textbf{13}, 3224 (1976); J. Bros
and U. Moschella, Rev. Math. Phys. \textbf{8}, 327 (1996); R. Bousso, A.
Maloney, and A. Strominger, Phys. Rev. D \textbf{65}, 104039 (2002).

\bibitem{Prud86} A.P. Prudnikov, Yu.A. Brychkov, and O.I. Marichev, \textit{%
Integrals and Series} (Gordon and Breach, New York, 1986), Vol. 2.

\bibitem{Rome02} A. Romeo and A.A. Saharian, J. Phys. A: Math. Gen. \textbf{%
35}, 1297 (2002).

\bibitem{Full03} S.A. Fulling, J. Phys. A: Math. Gen. \textbf{36}, 6857
(2003); K. A. Milton, J. Phys. A: Math. Gen. \textbf{37}, R209 (2004).

\bibitem{Saha01EMT} A.A. Saharian, Phys. Rev. D \textbf{69}, 085005 (2004).

\bibitem{Grah02} N. Graham, R.L. Jaffe, V. Khemani, M. Quandt, M. Scandurra,
and H. Weigel, Nucl. Phys. B \textbf{645}, 49 (2002); N. Graham, R.L. Jaffe,
and H. Weigel, Int. J. Mod. Phys. A \textbf{17}, 846 (2002); N. Graham, R.L.
Jaffe, V. Khemani, M. Quandt, O. Schr\"{o}der, and H. Weigel, Nucl. Phys. B
\textbf{677}, 379 (2004).

\bibitem{GradShtein} I.S. Gradshteyn and I.M. Ryzhik, \textit{Table of
Integrals, Series and Products} (Academic, New York, 1980).

\bibitem{Milt09} F.W.J. Olver, \textit{Asymptotics and Special Functions}
(Academic Press, New York, 1974); T.M. Dunster, SIAM J. Math. Anal. \textbf{%
21}, 995 (1990); K.A. Milton, J. Wagner, and K. Kirsten, Phys. Rev. D
\textbf{80}, 125028 (2009).
\end{thebibliography}
\end{document}